\documentclass[a4paper,12pt]{article}
\title{MSM_pseudo_CS}
\usepackage{amssymb, amsmath, amsthm, mathtools, color}
\usepackage{booktabs}       
\usepackage{multirow}
\usepackage{multicol}        
\usepackage{longtable}
\usepackage{array}
\usepackage{rotating}
\usepackage{float}
\usepackage{subfigure}
\usepackage{subfig}
\usepackage{epsfig}
\usepackage{url}
\usepackage{graphicx}
\usepackage{csquotes}
\usepackage{enumerate}
\usepackage{amsmath}
\usepackage{listings}
\usepackage{color}
\usepackage{physics}
\usepackage{listings}
\usepackage[all]{nowidow}
\usepackage{placeins}
\usepackage{comment}
\usepackage[title]{appendix}
\usepackage[numbers,super,sort&compress]{natbib}
\title{Pseudo-value regression of clustered multistate current status data with informative cluster sizes}
\author{Samuel Anyaso-Samuel${^1}$, Dipankar Bandyopadhyay${^2}$, Somnath Datta${^1}$ \\ \small ${^1}$Department of Biostatistics, University of Florida\\ \small ${^2}$Department of Biostatistics, Virginia Commonwealth University}
\date{}

\begin{document}
\maketitle

\begin{abstract}
Multistate current status (CS) data presents a more severe form of censoring due to the single observation of study participants transitioning through a sequence of well-defined disease states at random inspection times. Moreover, these data may be clustered within specified groups, and informativeness of the cluster sizes may arise due to the existing latent relationship between the transition outcomes and the cluster sizes. Failure to adjust for this informativeness may lead to a biased inference. Motivated by a clinical study of periodontal disease (PD), we propose an extension of the pseudo-value approach to estimate covariate effects on the state occupation probabilities (SOP) for these clustered multistate CS data with informative cluster or intra-cluster group sizes. In our approach, the proposed pseudo-value technique initially computes marginal estimators of the SOP utilizing nonparametric regression. Next, the estimating equations based on the corresponding pseudo-values are reweighted by functions of the cluster sizes to adjust for informativeness. We perform a variety of simulation studies to study the properties of our pseudo-value regression based on the nonparametric marginal estimators under different scenarios of informativeness. For illustration, the method is applied to the motivating PD dataset, which encapsulates the complex data-generation mechanism.
\end{abstract}

\section{Introduction}\label{sec: intro}
Multistate models \cite{andersen2002multi} are becoming increasingly popular in biomedical research for event history data analysis since these models are suitable for examining patient transitions between several well-defined disease states. The disease sequence defines the finite number of states in the model, and a patient is observed to occupy a known state at any given time throughout the observation period. Patients can move reversibly or irreversibly through a succession of intermediary states before transitioning to an absorbing state. Researchers often expect to measure the exact transition times between disease states for the population under study. However, they may be restricted to observing only the disease states occupied by the patients at their inspection times. Hence, only the current status of each patient concerning the known disease progression state is observed. Such an observation scheme induces a severe form of interval censoring (case-I), and the observed data are often called current status (CS) data. 

CS data arise naturally in several fields of biomedical research, including clinical trials, epidemiological studies, bioassay studies, and oral health. For example, in clinical cross-sectional studies in oral health, a patient's teeth-related outcomes may be observed at a single random inspection time, leading to CS data. Also, the CS disease outcomes of teeth recorded from the same patient (cluster) are expected to be correlated, due to shared behavioral and genetic patterns. Furthermore, the patients can have varying numbers of teeth, such that those who are more prone to the disease may have fewer teeth compared to those with good oral health \cite{mitani2021marginal}. This leads to the well-explored informative cluster size (ICS) scenario \cite{lan2017}, and failure to account for ICS will generally lead to biased estimates of the covariate effects and hypothesis tests with inflated case-I errors. 

\begin{figure}[h]
\centering
\includegraphics[width=.7\linewidth]{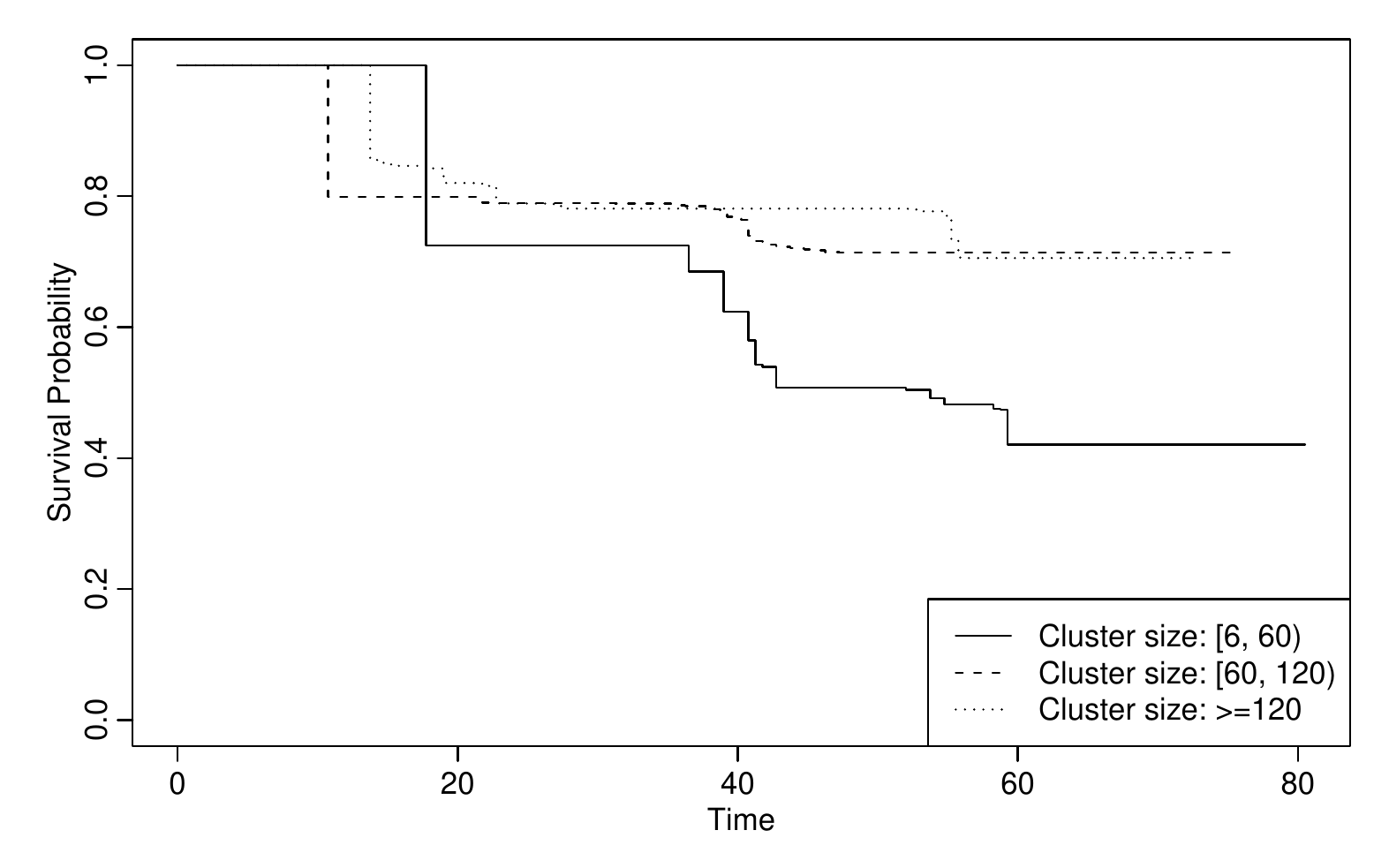}%
\caption{Nonparametric Turnbull estimates of (current status) tooth survival probability stratified by the number of inspected tooth-sites within a patient's mouth at the time of inspection, classified as patients with [6,60) inspected tooth sites, [60, 120) tooth sites, and $\geq 120$ tooth sites. The CS event of interest is moderate to severe PD at the inspection time.}
\label{fig: SCI_ics}
\end{figure}

To further motivate, consider the motivating cross-sectional study recording periodontal health among the Gullah-speaking African-American \cite{fernandes2009} diabetics (henceforth, the GAAD study, herein) at a given inspection time. Here, for each subject, the (correlated) periodontal disease (PD) status was measured at six sites per tooth excluding the molars, with the exact time of PD incidence unknown. PD disease progression is often postulated as an irreversible multistate model (MSM)\cite{mdala2014comparing} that comprises the following states: disease-free, slight PD, moderate PD, and severe PD (see Section \ref{sec: application} for a detailed description of the states). Under a traditional survival analysis setup where the event of interest is moderate to severe PD at the time of inspection, Figure \ref{fig: SCI_ics} presents the nonparametric Turnbull estimates of tooth survival curves, stratified by the number of tooth sites (168 sites in total, if all teeth were present). The plots reveal that patients with fewer inspected teeth (sites) had worse tooth survival, compared to those with more teeth, indicating features of ICS in this dataset. 

Datasets, such as the GAAD, are commonplace in most dental and oral epidemiological studies, with the following unique characteristics: correlated outcomes, ICS, and multistate event times, subject to case-I interval censoring. The objective of this paper is to conduct a prudent cross-sectional risk assessment on the probability of occupying a particular PD state at a given inspection time. Regression analysis for CS data has been primarily discussed in the context of a traditional survival setup (two-state model) where parametric or semiparametric methods are used for inference \cite{jewell2003, shiboski1998}. \textcolor{black}{In the context of multivariate current-status data, Sun and colleagues \cite{sun2006,chen2009,li2022new}  proposed regression methods based on additive hazards or proportional hazards frailty models. Even if one extends these methods to CS data from a MSM perspective, they may not provide direct estimates of the covariate effects on the temporal functions (e.g. state occupation probabilities) of the MSM.}

Regression methods for CS data from an MSM are limited in the literature. Recently, Lan et al \cite{lan2017} developed a nonparametric regression model applicable to clustered-correlated CS data from an MSM. Even though their model can account for ICS, it accommodates only a single covariate. To the best of our knowledge, there are no available methods or software for multivariable regression based on the CS data from MSMs. Pseudo-value regression \cite{andersen2003} is a robust alternative for measuring the association between covariates and temporal functions of an MSM. Originally developed in the context of uncorrelated time-to-event data that is subject to right-censoring, the regression analysis is typically conducted in two steps; (i) estimation of the marginal functions of the MSM, and (ii) fitting the estimating equations that are based on the (response) pseudo-values. In the case where the transition times are clustered, and the cluster size is informative, implementing the pseudo-value regression is not straightforward since one may need to adjust for ICS in the two aforementioned steps. 

In this article, we study the pseudo-value approach for cluster-correlated CS data with ICS and describe appropriate guidelines for its application in quantifying covariate effects on the state occupation probability (SOP) of an MSM. In the first step of the analysis, we estimate the marginal SOP using a combination of nonparametric regression theory and the product limit estimation as described by Datta and Sundaram \cite{datta2006}. We study the properties of the jackknife pseudo-values that are based on nonparametric regression estimators and provide theoretical justifications for the subsequent regression analysis. For estimating the covariate effects in the second step, we formulate the estimating equations with weights that are inverse of the cluster sizes to adjust for informative cluster size. By reweighting only the estimating equations in the pseudo-value approach, we adopt the recommendation of Anyaso-Samuel and Datta \cite{anyasosamuel2022} who investigated the pseudo-value regression in the context of right censored multistate time-to-event data with ICS.  

The rest of the article is as follows. In Section \ref{sec: marg_est}, we formulate a nonparametric marginal estimator of the SOP for the multistate cluster-correlated CS data with ICS. In Section \ref{sec: pseudo}, we describe the implementation of the pseudo-value regression. Section \ref{sec: sim_CS} contains simulation studies to evaluate the empirical performance of the pseudo-value approach under various data generation scenarios. In Section \ref{sec: application}, we apply the pseudo-value regression approach to data from the GAAD study. The article concludes with a discussion in Section \ref{sec: discussion}.  Theoretical arguments, and additional simulation results are relegated to Appendix A, and Web-supplements A--C, respectively.

\section{Nonparametric estimators for a multistate model}\label{sec: marg_est}
\subsection{Notation and Convention}\label{sec: notation}
Let $\mathcal{S}= \{1,2,\ldots, Q\}$ be the finite state space for the underlying multistate system with a directed tracking structure. Let $X_{ij}(t) \in \mathcal{S}$ denote the state occupied at time $t \geq 0$ and $C_{ij}$ denote the random inspection time for individual $j = 1,\ldots,n_i$ in cluster $i=1,\ldots,m$. Define $n=\sum_{i=1}^m n_i$. The observed CS data for such an individual is represented as the triple $\{C_{ij}, X_{ij}(C_{ij}),\mathbf{Z}_{ij}\}$ where $C_{ij}$ is independent of the multistate process $\{X_{ij}(t), t \geq 0 \}$ given the $p$-dimensional covariate vector $\mathbf{Z}_{ij}$. The observations within the clusters may be correlated, while the observations across distinct clusters are independent. 

Our central objective is to model the SOPs which are defined as $\pi_\ell(t) = \text{Pr}\{X(t) = \ell\} = \sum_{\ell' \in \mathcal{S}} \pi_{\ell'}(0)P_{\ell' \ell}(0,t),$ where $\pi_{\ell'}(0)$ is the initial occupation probability for state $\ell'$, and $P_{\ell' \ell}(s,t)$ is the $\ell \ell'$th element of the transition probability matrix $\boldsymbol{P}(s,t)$. The counting process denoted by $N_{\ell \ell'}(t)$ and the at-risk process denoted by $M_{\ell}(t)$ is important functionals for estimating $\pi_\ell(t)$ and other temporal functions of the multistate model. Here, $N_{\ell \ell'}(t)$ counts the number of transitions from state $\ell$ to $\ell'$ in the time interval $[0, t)$, and $M_{\ell}(t)$ represents the number of individuals at-risk of transitioning out of state $\ell$ at time $t$.

\subsection{Marginal estimators for the clustered data}\label{sec: unclust_data}
Datta and Sundaram \cite{datta2006} proposed estimators of $N_{\ell \ell'}(t)$ and $M_{\ell}(t)$ using the theory of nonparametric regression. We briefly describe the development of the resulting estimators under the setting where the data are cluster-correlated. Let $U_{\ell \ell'}$ denote the (unobserved) transition time at which an individual transition from $\ell$ to $\ell'$. Then, $N_{\ell \ell'}(t)$ and $M_\ell(t)$ can be expressed as follows
\begin{align*}
    N_{\ell \ell'}(t) &= \sum^{m}_{i=1}\sum^{n_i}_{j=1} \mathcal{I}(U_{ij,\ell \ell'} \leq t), \quad \text{ and }\\
    M_\ell(t) &= \sum^{m}_{i=1}\sum^{n_i}_{j=1} \mathcal{I}(X_{ij}(t-) = \ell),
\end{align*} 
where $\mathcal{I}(\cdot)$ denotes the indicator function and $X(t-)$ is the state occupied just before time $t$. By the law of large numbers,
\begin{align*}
    n^{-1}N_{\ell \ell'}(t) \xrightarrow{P} \text{Pr}(U_{\ell \ell'} \leq t),
\end{align*} 
for any $t \geq 0$. Since the inspection time $C$ is independent of the multistate process, we can write 
\begin{align*}
    \text{Pr}(U_{\ell \ell'} \leq t) = \mathbb{E}(\mathcal{I}(U_{\ell \ell'} \leq C)\ |\ C = t),
\end{align*} 
where $\mathcal{I}(U_{\ell \ell'} \leq C)$ is the indicator of the event that the transition from state $\ell$ to $\ell'$ has taken place by time $C$. Note that $\text{Pr}(U_{\ell \ell'} \leq t)$ is monotonic in $t$, Datta and Sundaram \cite{datta2006} suggested a two-step approach for estimating $n^{-1}N_{\ell \ell'}(t)$. First, perform isotonic regression based on the pairs $(C_{ij},\ \mathcal{I}(U_{ij,\ell \ell'} \leq C_{ij}))$ using the pooled adjacent violators (PAV) algorithm \cite{barlow1972}, and let $\widehat{N}^P_{\ell \ell'}(t)$ denote the solution to the weighted sum of squares with weights $w_{ij} \in \{1, \frac{1}{n_i}\}$. Secondly, perform kernel smoothing \cite{mukerjee1988, nadaraya1964, watson1964} with a log-concave density, $K > 0$, to remove the long flat parts of $\widehat{N}^P_{\ell \ell'}(t)$ obtained from the first step. Note that, $\mathcal{I}(U_{\ell \ell'} \leq C)$ can be obtained from the observed CS data. The final estimator of $N_{\ell \ell'}(t)$ is 
\begin{align} 
    \widehat{N}_{\ell \ell'}(t) =  \sum^{m}_{i=1}\sum^{n_i}_{j=1} \left\{ w_{ij} \times \widehat{N}^P_{\ell \ell'}(t) \times \frac{ K_h(C_{ij} - t)}{\sum^{m}_{i=1}\sum^{n_i}_{j=1} K_h(C_{ij} - t)} \right\},
    \label{eqn: CPest}
\end{align} 
where $K_h(\cdot) = h^{-1}K(\cdot/h)$ with a bandwidth sequence $0 < h = h(n) \downarrow 0$ chosen by the criteria specified by Wand and Jones \cite{wand1994}.
Also, given that $\text{Pr}(X(t -) = \ell)$ is the probability limit of $n^{-1}M_{\ell}(t)$, Datta and Sundaram \cite{datta2006} utilized the kernel smoothing approach for estimating $M_{\ell}(t)$. Thus,
\begin{align} 
    \widehat{M}_{\ell}(t) = \sum^{m}_{i=1}\sum^{n_i}_{j=1} \left\{w_{ij} \times \mathcal{I}(X_{ij}(C_{ij}) = \ell) \times \frac{K_h(C_{ij} - t)}{n^{-1}\sum^{m}_{i=1}\sum^{n_i}_{j=1} K_h(C_{ij} - t)} \right\},
    \label{eqn: ARest}
\end{align} 
where $X_{ij}(C_{ij} -) = X_{ij}(C_{ij})$ with probability 1, by assumption. Since the at-risk set out of the initial state is monotonic in $t$, one may be able to add the PAV step for estimating the risk set out of this state. The asymptotic properties of these estimators can be established following the arguments outlined in Datta and Sundaram \cite{datta2006}. 

The estimators, $\widehat{N}_{\ell \ell'}(t)$ and $\widehat{M}_{\ell}(t)$ are used to estimate the integrated transition hazard, ${\boldsymbol{A}}(t)$. Let $\widehat{\boldsymbol{A}}(t)$ denote the estimator of ${\boldsymbol{A}}(t)$; its elements are expressed by
\begin{align} 
        \widehat{A}_{\ell\ell'}(t) = 
    \begin{dcases}
    \int^t_0 \mathcal{I} \Big(\widehat{M}_{\ell}(u)>0 \Big) \widehat{M}_{\ell}(u)^{-1} \text{d}\widehat{N}_{\ell\ell'}(u) & \ell \neq \ell',\\
    - \sum_{\ell' \neq \ell} \widehat{A}_{\ell\ell'}(t) & \ell = \ell'.
    \end{dcases}
    \label{eqn: NelsonAalen}
\end{align} 
Further, the marginal estimator of the transition probability matrix, $\widehat{\boldsymbol{P}}(s,t)$, is obtained by the product limit of $\widehat{A}_{\ell\ell'}(t)$,
\begin{align} 
    \widehat{\boldsymbol{P}}(s,t) = \prod_{(s,t]} \{\boldsymbol{I} + \text{d}\widehat{\boldsymbol{A}}(u) \},
    \label{eqn2: transProb}    
\end{align} 
where $\boldsymbol{I}$ denotes the $Q \times Q$ identity matrix. Finally, the marginal estimator of the SOP is defined by 
\begin{align} 
    \widehat{\pi}_\ell(t) = \sum^Q_{\ell'=1} \widehat{\pi}_{\ell'}(0) \widehat{P}_{\ell' \ell}(0, t),
   \label{eqn: stocc}
\end{align} 
where $\widehat{\pi}_{\ell'}(0) = \{\widehat{M}_{\ell'} (0+)\}/\{\sum^Q_{\ell=1} \widehat{M}_\ell (0+)\}$, and $\widehat{P}_{\ell' \ell}(0, t)$ is the $\ell' \ell$th element of $\widehat{\boldsymbol{P}}(0,t)$. Andersen et al \cite{andersen1993} explained the relationship among (\ref{eqn: NelsonAalen}) - (\ref{eqn: stocc}), while Datta and Satten \cite{datta2001} showed the validity of (\ref{eqn: stocc}) for a possible non-Markov model.

We formulate the marginal estimators with weights, $w_{ij} \in \{1, \frac{1}{n_i}\}$. In the scenario where the cluster sizes are informative, inverse cluster reweighting has been suggested in the literature to arrive at correct marginal inference since the marginal estimators formulated with $w_{ij} = 1$ may lead to biased estimates and inflated type-I errors. By inverse cluster size reweighting, we randomly select a typical patient from a typical cluster thereby allowing equal total contribution from all clusters in the estimation procedure.

The marginal estimator of the SOP described in this section follows the theory of nonparametric regression. If covariates are available, researchers are often interested in quantifying the effect of such covariates on the temporal function of interest. Following a similar nonparametric regression theory presented in this section, Lan et al \cite{lan2017} proposed a conditional estimator of the SOP given a single covariate. Besides the smoothing parameter (bandwidth) $h$ required in (\ref{eqn: CPest}) and (\ref{eqn: ARest}), their method requires the specification of another smoothing parameter that depends on the single covariate for computing the weights need for the isotonic regression of $\mathcal{I}(U_{\ell \ell'} \leq C)$ on $C$.  If the inference is desired for multiple covariates, the method of Lan et al becomes inapplicable. Besides, a fully nonparametric regression modeling approach will suffer from the curse of dimensionality due to the need to specify a smoothing parameter for each covariate. The pseudo-value regression presents a flexible approach for estimating the association between multiple covariates and the quantity $\pi_\ell(t)$, without requiring additional smoothing parameters.

\section{Pseudo-value regression}\label{sec: pseudo}
Previously, we defined the marginal estimator for the mean-valued quantity $\pi_\ell(t) = \mathbb{E}\{\mathcal{I}(X(t) = \ell)\} = \mathbb{P}\{X(t) = \ell\}$. Our interests lie in estimating the effects of covariates $\mathbf{Z}$ on the SOP. In other words, we seek to model the quantity $\mathbb{E}(\mathcal{I}\{X(t) = \ell\} \ |\ \mathbf{Z}) = \mathbb{P}\{X(t) = \ell \ |\ \mathbf{Z}\}$. The pseudo-value approach was proposed by Andersen et al. \cite{andersen2003} to conduct regression analysis when modeling temporal functions of a multistate model, for example, the SOP. Consider the setting where the CS times $\{C_{i},\ i=1,...,n\}$ are uncorrelated. Given an (approximately) unbiased estimator $\widehat{\pi}_\ell(t)$ of the SOP, the pseudo-values are defined by
\begin{align*} 
    Y_{i}(t) = n \cdot \widehat{\pi}_\ell(t) - (n-1) \widehat{\pi}_{\ell,-i}(t),
\end{align*} 
where $\widehat{\pi}_{\ell,-i}(t)$ is obtained by omitting the $i$th subject. For the uncorrelated CS times, (\ref{eqn: CPest}) and (\ref{eqn: ARest}) will be formulated with a single sum over $n$ observations, and $w_i = 1$. Under the CS setting, $Y_{i}(t)$, just like the jackknife, will denote the relative contribution of the $i$th subject in an approximately linear relationship leading to the marginal estimator. The pseudo-values are then used as the responses in a regression model to quantify the effects of the covariates on the SOP.

\subsection{Properties of the pseudo-values} \label{ssec:property} 
\begin{figure}[h]
\centering
\includegraphics[width=.7\linewidth]{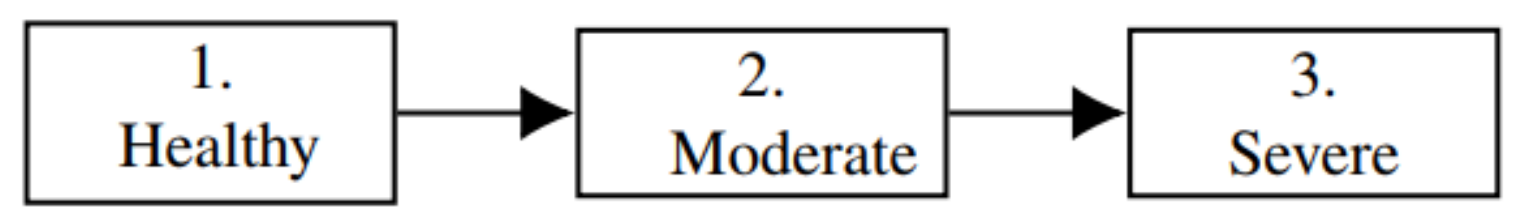}%
\caption{The 3-state tracking model.} \label{fig: tracking}
\end{figure}

\begin{figure}[ht]
\hspace{-0.3cm}
\includegraphics[scale=0.95]{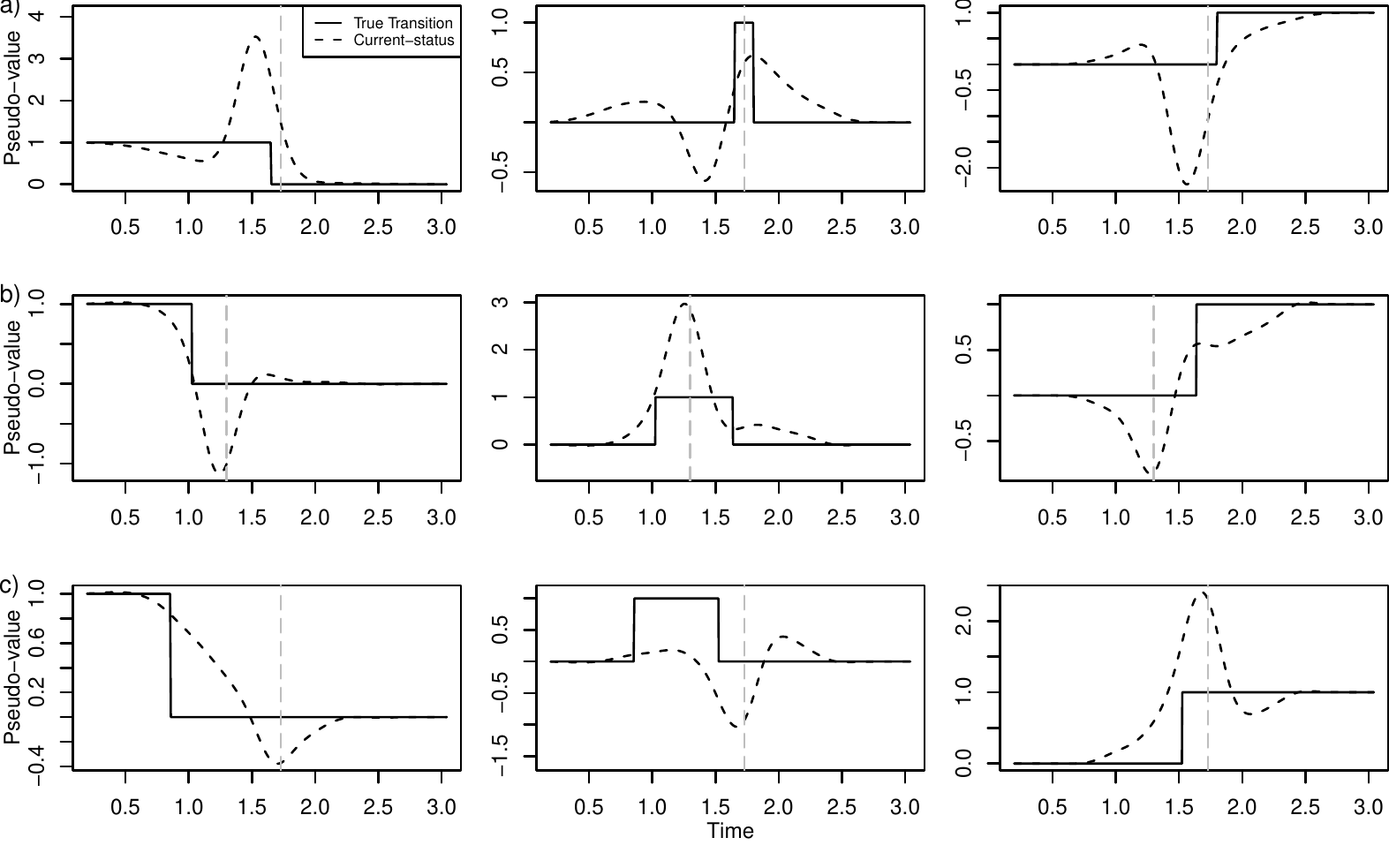}%
\caption{Pseudo-values based on true transition times (solid lines) and current-status data (dashed lines) for the occupation probability of initial (left), transient (center), and absorbing (right) state of a three-state tracking model (see Figure \ref{fig: tracking}). The plots (top, middle, bottom) of the pseudo-values are shown for three different patients that were inspected while occupying the three respective states of the tracking model. The broken vertical line in each plot indicates the inspection time for a given patient.}
\label{fig: pseudo}
\end{figure}

For a simulated cohort of $n=1,000$ patients whose disease evolution begins at the initial state of a three-state tracking model (see Figure \ref{fig: tracking}), the top, middle, and bottom panels of Figure \ref{fig: pseudo} show the pseudo-values for three different patients, namely (a), (b), and (c) who were occupying states 1, 2, and 3, respectively at their inspection time. The pseudo-values for these patients are based on the entire cohort and were computed at all observed inspection times. The solid lines in Figure \ref{fig: pseudo} represent the pseudo-values calculated using the empirical estimator of the SOP with the exact transition times, while the broken lines represent the pseudo-values for the observed CS data based on the product-limit estimator. Under the analysis of exact transition times, the pseudo-values indicate whether or not patient $i$ is in a particular state at time $t$, that is, the pseudo-value is 1 when the patient occupies the state at time $t$ and 0, otherwise. For the CS data, let $Y^{[1]}$ (left column of Figure \ref{fig: pseudo}), $Y^{[2]}$ (center), and $Y^{[3]}$ (right) denote the pseudo-values corresponding to $\widehat{\pi}_1$, $\widehat{\pi}_2$, and $\widehat{\pi}_3$, respectively. Notice that $Y^{[1]}$ is initially one since we assume that the disease trajectory of every patient begins at the initial state. For the patient who was inspected while occupying state 1, as time passes, $Y^{[1]}$ increases and has a turning point around the inspection time. However, for patients who were in other states at their inspection times, $Y^{[1]}$ mostly decreases and has a turning point around the inspection time. Yet, $Y^{[1]}$ gradually reduces to 0 since all patients will eventually transition out of state 1. Also, consider the pseudo-values for the absorbing state; for the patient inspected while occupying this state, $Y^{[3]}$ increases as time passes but has a turning point at inspection time. The converse is the case for patients who were not in state 3 when they were inspected, albeit, $Y^{[3]}$ is eventually 1 since all patients are likely to transition permanently to this absorbing state. For state 2, which is the transient state, the pseudo-value for each patient is initially 0.  As time progresses, $Y^{[2]}$ increases for the patient inspected while occupying state 2 with a turning point at the inspection time. Ultimately, $Y^{[2]}$ returns to 0 for each patient since they will eventually transition to state 3. In general, the pseudo-values for the CS data correspond to the smoothed version of their empirical counterparts that are based on the complete data.

\subsection{A rationale for the pseudo-value approach}\label{ssec:rationale} 

Again, consider the scenario where the observed data $\{(C_{i}, X_{i}(C_{i}),\mathbf{Z}_{i}),\ i=1,\ldots,n\}$ are uncorrelated, and the SOP is the mean-valued function of interest. Following the idea of nonparametric smoothing, consider the na\"ive (empirical) estimator 
\begin{align*}
    \widetilde{\pi}_{\ell}(t) = \frac{1}{2hn} \sum_{i=1}^{n} \mathcal{I} \{X_{i}(C_{i}) = \ell,\ t-h < C_i < t+h\}
\end{align*} 
where $0<h=h(n) \downarrow 0$ is a user-defined bandwidth sequence. The Aalen-Johansen-type estimator $\widehat{\pi}_{\ell}(t)$ given in (\ref{eqn: stocc}) is a refined version of the natural estimator $\widetilde{\pi}_{\ell}(t)$, and we explain the utility of the pseudo-value regression in the context of the CS data using $\widetilde{\pi}_{\ell}(t)$. Let $\widetilde{\pi}_{\ell,-i}(t)$ denote the empirical estimator based on omitting the $i$th observation, the corresponding pseudo-value for the $i$th subject is given by
\begin{align}
    Y_i(t) = \frac{1}{2h} \mathcal{I} \{X_{i}(C_{i}) = \ell,\ t-h < C_i < t+h\}.
    \label{eqn: pseudo1}
\end{align} 
Further, regressing $Y_i(t)$ on $\mathbf{Z}_{i}$ leads to
\begin{align*} \begin{aligned}
    \mathbb{E}\{Y_i(t) \ |\ \mathbf{z} \} &= \frac{1}{2h} \mathbb{P}\{X_{i}(C_{i}) = \ell,\ t-h < C_i < t+h \ | \ \mathbf{Z}_{i}=\mathbf{z} \}\\
    &\approx \mathbb{P}\{X_i(t) = \ell \ | \ \mathbf{Z}_{i}=\mathbf{z} \}
\end{aligned} \end{align*} 
for small $h$. Therefore, the regression of the jackknife pseudo-values on the covariates will lead to correct approximate covariate inference on the probability of occupying a given state. For the scenario where the data are clustered, an extension of these arguments is presented in Appendix \ref{app: appenA}.

\subsection{Estimation of the covariate effects}\label{ssec:covariate}

Now, we return to the original scenario where the observed data $\{C_{ij}, X_{ij}(C_{ij}),\mathbf{Z}_{ij}\}$ are cluster-correlated. We are interested in estimating the measure of association between available covariates and the occupation probabilities for a typical patient in a typical cluster. The pseudo-values may be used as responses in a standard regression model to perform the covariate inference.

Let $\boldsymbol{Y}_{ij} = \{Y_{ij}(t_1),\ldots,Y_{ij}(t_r)\}$ denote the vector of pseudo-values calculated at time points $\{t_1,\ldots,t_r\}$. Define $g(\pi(t_k|\mathbf{Z}_{ij})) = \alpha_k + \boldsymbol{\zeta} \mathbf{Z}_{ij} = \boldsymbol{\beta} \mathbf{Z}^*_{ijk}$, where $\mathbf{Z}^*_{ijk} = (\mathcal{I}(t_l = t_k), l=1,\hdots,r; \mathbf{Z}_{ij})$. This allows a different intercept for each of the $k=1,\hdots,r$ time points, and $\boldsymbol{\zeta}$ describes the covariate effects on the mean changes in the $\boldsymbol{Y}_{ij}$. Moreover, one can study the time-varying effects of the covariates by modeling the interaction between the covariates and the time points in a regression model. \textcolor{black}{In the setting where the event times are uncorrelated, previous articles \cite{klein2005, andersen2010} have studied the sensitivity to the choice of the number of time points for pseudo-value based inference. The recommendation is to compute the pseudo-values at five-time points, at least, equally spread on the event/time scale.}

\textcolor{black}{To estimate $\boldsymbol{\beta}$, we solve the following estimating equation
\begin{align} 
\sum^{m}_{i=1} \sum^{n_i}_{j=1} w_{ij} \cdot \boldsymbol{U}_{ij}(\boldsymbol{\beta}, \alpha) = \boldsymbol{0},
\label{eqn: estimating}
\end{align} 
where 
$$
\boldsymbol{U}_{ij}(\boldsymbol{\beta}, \alpha) = \boldsymbol{D}_{ij}^T\boldsymbol{V}^{-1}_{ij} (\boldsymbol{Y}_{ij}  - g^{-1}(\boldsymbol{\beta} \mathbf{Z}^*_{ij})), 
$$
$\boldsymbol{D}_{ij} =\Big(\frac{\partial g^{-1}(\boldsymbol{\beta} \mathbf{Z}^*_{ij})}{\partial \boldsymbol{\beta}} \Big)$ and $\boldsymbol{V}_{ij} = \sigma^2 \boldsymbol{R}_{ij}(\alpha)$ is the working-covariance matrix. Further, we estimate the variance of $\widehat{\boldsymbol{\beta}}$ using the sandwich estimator given by
\begin{align}
    \widehat{\text{Var}}(\widehat{\boldsymbol{\beta}}) = \boldsymbol{B}^{-1} \boldsymbol{M} \boldsymbol{B}^{-1}
    \label{eqn: sandwich}
\end{align}
where 
\begin{align*}
    \boldsymbol{B} &= \sum^{m}_{i=1} \sum^{n_i}_{j=1} w_{ij} \cdot \boldsymbol{D}_{ij}^T \widehat{\boldsymbol{V}}^{-1}_{ij} \boldsymbol{D} \\
    \boldsymbol{M} &= \sum^{m}_{i=1} \left \{\left(\sum^{n_i}_{j=1} w_{ij} \cdot \boldsymbol{U}_{ij}(\widehat{\boldsymbol{\beta}}, \widehat{\alpha}) \right) \left(\sum^{n_i}_{j=1} w_{ij} \cdot \boldsymbol{U}_{ij}(\widehat{\boldsymbol{\beta}}, \widehat{\alpha})\right)^T \right \}
\end{align*}
with $\widehat{\boldsymbol{V}}_{ij} = \widehat{\sigma}^2 \boldsymbol{R}_{ij}(\widehat{\alpha})$, and $\widehat{\sigma}$ and $\widehat{\alpha}$ are consistent estimates of the $\sigma$ and $\alpha$, respectively (see Sections 2.1 \& 2.3 of Wang et al \cite{wang2011}). Here, we utilize the weights $w_{ij} \in \{1, \frac{1}{n_i}\}$. Wang et al \cite{wang2011} discuss the properties of (\ref{eqn: estimating}) for both choices of $w_{ij}$, and varying choices of the working-covariance matrix for the analysis of longitudinal clustered data.}

Generally, under the scenario where the cluster sizes are informative, the traditional GEE \cite{liang1986} (with $w_{ij}=1$) leads to biased inference since it assumes that the outcomes are independent of the cluster sizes. Williamson et al \cite{williamson2003} proposed the cluster-weighted GEE (CWGEE) that adjusts for ICS by reweighting the standard GEE with the inverse of the cluster size. For a regression analysis with the CWGEE, all clusters from the data contribute equally to the analysis, while for the GEE model, larger clusters contribute more than the smaller clusters. If the cluster sizes are informative, both marginal regression models will yield different inferential conclusions. Under the context of the pseudo-value regression, we provide theoretical justifications for the inverse cluster size reweighting for marginal analysis in Appendix \ref{app: appenA}.

Situations also exist in multilevel designs where there are known groups within the clusters, and a latent relationship exists between the number of units with the same group membership and the observed outcomes for such groups. This phenomenon is often termed an informative intra-cluster group (ICG) size \cite{dutta2016}. Let $G_{ij} \in \{0,1\}$ denote the group membership of patient $j$ within cluster $i$, and $n_{G_{ij}}$ denotes the number of observations in cluster $i$ with the same group membership as patient $j$. To adjust for informative ICG sizes in marginal analysis, Huang and Leroux \cite{huang2011} proposed the doubly-weighted estimating equations (DWGEE) that estimate the regression parameters by reweighting the contribution from a patient $j$ within a given cluster $i$ by $w_{ij} = \frac{1}{n_{G_{ij}}}$. In Web supplement \textcolor{red}{C}, we will return to the analysis of cluster-correlated CS data under ICG via simulations. 

Apparently, for the pseudo-value-based regression involving clustered data with ICS, the adjustment for ICS by the inverse cluster size reweighting can be implemented in the marginal estimation of the SOP and while solving the estimating equations. For the case with clustered event times subject to right-censoring, Anyaso-Samuel and Datta \cite{anyasosamuel2022} showed that valid covariate inference is obtained by primarily adjusting for ICS when fitting the estimating equations that are based on correct pseudo-values. In the rest of this article, we calculate the pseudo-values, $Y_{ij}(t)$ using the unweighted ($w_{ij}=1$) marginal estimators of the SOP.

\section{Simulation Study}\label{sec: sim_CS}
In this section, we conduct simulation studies to study the finite sample performance of our estimators. Our data generation scheme closely mimics the GAAD data setting introduced earlier. In particular, we consider a 3-state tracking model (shown in Figure \ref{fig: tracking}), where states $1-3$ denote healthy, moderate, and severe conditions of the disease, respectively. Our primary goal is to evaluate the implementation of the proposed pseudo-value approach for inference when analyzing cluster-correlated CS data with ICS. In addition, we also compare methods for adjustment of ICS in fitting the pseudo-values to the covariates.

Our simulated data comprise $(C_{ij}, S(C_{ij}), \mathbf{Z}_{ij})$, $i=1,\ldots,m$ and $j=1,\ldots,n_i$. We simulate the cluster-correlated exit times from state 1, denoted by $T_{1,ij}$, via a lognormal model given by
\begin{align} 
    \log(T_{1 \cdot,ij}) = \delta_1 Z_{1,i} + \delta_2 Z_{2,ij} +  \nu_i + \sigma\varepsilon_{ij}
    \label{eqn: lognormModel}
\end{align} 
where $Z_{1,i}$ is a cluster-level binary covariate, $Z_{2,ij} \sim \text{N}(0, 0.15)$ is a subject-level continuous covariate, $\nu_i \sim \text{N}(0, 0.25)$ is a cluster-level random effect, and $\varepsilon_{ij} \sim \text{N}(0,\ 1)$ is the subject-level random error. In the PD example, $\nu_i$ captures the between-teeth correlation induced by shared behavioral and genetic markers. The true exit times from state 2, denoted by $T_{2,ij}$, are obtained by $T_{2,ij} = D^{-1} [D(T_{1,ij}) + R_2 \{1 - D(T_{1,ij}) \} ]$ where $D(\cdot)$ and $D^{-1}(\cdot)$ respectively denotes the distribution function and quantile function of the lognormal distribution with parameters $\boldsymbol{\delta}^T\boldsymbol{Z}$ and $\sigma$, respectively, and $R_2$ is generated from the standard uniform distribution. We simulate the clustered inspection times from a Weibull distribution with parameters, $\eta = 3$ (shape), and $\tau =5$ (scale). The cluster sizes are generated from the informative and non-informative scenarios. For the ICS, we simulate $n_i \sim \text{Poisson}\{\exp(1.5+3\nu_i -3Z_{1, i})\}+2$, such that the dependence of $n_i$ on the cluster-specific binary covariate and the random effect induces ICS. For the non-ICS situation, we simulate $n_i \sim \text{Poisson}(20)$. For each setting, we consider two choices for the number of clusters; $m=30$ and $200$, and we set $(\delta_1,\delta_2) = (0.25,\ 0.8)$ and $\sigma = 0.3$. Unless stated otherwise, we utilize an identity link for the pseudo-value regression. All simulation results presented using the various methods are evaluated with 1,000 Monte Carlo iterates.


\subsection{Estimation accuracy for a single continuous covariate}
First, we compare the performance of our pseudo-value approach with the method proposed by Lan et al \cite{lan2017} in estimating the occupation probability conditioned on the continuous covariate, $Z_2$. Specifically, we aim to estimate $\pi_\ell(t|z) = \mathbb{P}\{X(t) = \ell \ |\ Z_2=z\}$. We evaluate this quantity at $t=3.0$. Note that the nonparametric conditional estimators due to Lan et al are reweighted by the inverse of the cluster size. On the other hand, estimates from the pseudo-value approach are based on the regression analysis of the pseudo-value responses on $Z_2$; the pseudo-values are obtained from the nonparametric marginal estimator of the SOP. Further, we compare the performance of the regression model based on the CWGEE and the GEE in estimating $\pi_\ell(t|z)$. Although the pseudo-value regression with an identity link will lead to reliable tests of the effect of $Z_2$ on the occupation probability, this approach may occasionally yield estimates of $\pi_\ell(t|z)$ that lie outside the interval [0, 1]. To get range preserving estimates of $\pi_\ell(t|z)$, howbeit, at the cost of a slight increase in estimation bias, one may transform the pseudo-value responses $Y_{ij}(t)$ as binary and utilize a logit mean-link function in the estimating equation. Under this specification, one may set the responses to be $1$ if $Y_{ij}(t) > 0.5$ and 0, if otherwise. If the primary interest lies in estimating and testing the effect of the covariate on the SOP, the binary transformation will be unnecessary.

We assess the performance of the estimators at the first, second, and third quartiles of $Z_2$ which we denote by $Z_2^{Q_1}$, $Z_2^{Q_2}$, and $Z_2^{Q_3}$, respectively. The target quantities for each state are given by 
\begin{align}
\begin{aligned}
    \widetilde{\pi}_1(t|z) &= \frac{\sum^m_{i=1} \frac{1}{n_i} \sum^{n_i}_{j=1} \mathcal{I}(T_{1,ij} > t) \phi (\frac{Z_{2,ij} - z}{h})}{\sum^m_{i=1} \frac{1}{n_i} \sum^{n_i}_{j=1} \phi (\frac{Z_{2,ij} - z}{h})} \\
    \widetilde{\pi}_2(t|z) &= \frac{\sum^m_{i=1} \frac{1}{n_i} \sum^{n_i}_{j=1} \mathcal{I}(T_{1,ij} < t < T_{2, ij}) \phi (\frac{Z_{2,ij} - z}{h})}{\sum^m_{i=1} \frac{1}{n_i} \sum^{n_i}_{j=1} \phi (\frac{Z_{2,ij} - z}{h})} \\
    \widetilde{\pi}_3(t|z) &= \frac{\sum^m_{i=1} \frac{1}{n_i} \sum^{n_i}_{j=1} \mathcal{I}(T_{2,ij} < t) \phi (\frac{Z_{2,ij} - z}{h})}{\sum^m_{i=1} \frac{1}{n_i} \sum^{n_i}_{j=1} \phi (\frac{Z_{2,ij} - z}{h})}
\end{aligned}
\end{align} 
where $\phi(\cdot)$ is standard normal kernel with bandwidth $h$. The targeted quantities are computed using the true exit times $T_{1,ij}$ and $T_{2,ij}$ simulated from a single large sample of $m=1000$ clusters.

\begin{table}
\caption{Bias and Monte Carlo (MC) error (i.e., MC standard deviation of the estimates, in parentheses) of the estimators of the conditional occupation probabilities for the simulation scenario where the cluster sizes are informative. Also, idty = identity link, logit = logit link.}
\label{tab: cond_est}
\resizebox{1.0\textwidth}{!}{%
\begin{tabular}{rlllllllllll}
\toprule
\multicolumn{3}{c}{ } & \multicolumn{3}{c}{Healthy} & \multicolumn{3}{c}{Moderate} & \multicolumn{3}{c}{Severe} \\
\cmidrule(l{3pt}r{3pt}){4-6} \cmidrule(l{3pt}r{3pt}){7-9} \cmidrule(l{3pt}r{3pt}){10-12}
$m$ & Model & Link & $Z_2^{Q_1}$ & $Z_2^{Q_2}$ & $Z_2^{Q_3}$ & $Z_2^{Q_1}$ & $Z_2^{Q_2}$ & $Z_2^{Q_3}$ & $Z_2^{Q_1}$ & $Z_2^{Q_2}$ & $Z_2^{Q_3}$\\
\midrule
 &  &  & 0.2240 & -0.0078 & -0.3359 & 0.2491 & 0.0032 & -0.2742 & 0.2585 & 0.0017 & -0.1844\\

 &  & \multirow{-2}{*}{idty} & (0.1309) & (0.0597) & (0.1055) & (0.1455) & (0.0664) & (0.1172) & (0.1600) & (0.0730) & (0.1289)\\

 &  &  & 0.3449 & -0.0290 & -0.3165 & 0.3199 & -0.0577 & -0.2916 & 0.2778 & -0.0958 & -0.2359\\

 & \multirow{-4}{*}{GEE} & \multirow{-2}{*}{logit} & (0.2128) & (0.0497) & (0.1114) & (0.2271) & (0.0481) & (0.1140) & (0.2399) & (0.0460) & (0.1158)\\

 &  &  & 0.0281 & 0.0169 & -0.1649 & 0.0314 & 0.0306 & -0.0842 & 0.0191 & 0.0319 & 0.0246\\

 &  & \multirow{-2}{*}{idty} & (0.1075) & (0.0717) & (0.1172) & (0.1195) & (0.0797) & (0.1302) & (0.1314) & (0.0877) & (0.1432)\\

 &  &  & 0.2512 & -0.0197 & -0.2218 & 0.2226 & -0.0487 & -0.1966 & 0.1779 & -0.0871 & -0.1418\\

 & \multirow{-4}{*}{CWGEE} & \multirow{-2}{*}{logit} & (0.1845) & (0.0538) & (0.1007) & (0.1999) & (0.0529) & (0.1072) & (0.2142) & (0.0514) & (0.1128)\\

 &  &  & 0.0282 & 0.0445 & -0.0727 & 0.0299 & 0.0442 & -0.0741 & 0.0246 & 0.0302 & -0.0548\\

\multirow{-10}{*}{\raggedleft\arraybackslash 30} & \multirow{-2}{*}{Lan et al.} & \multirow{-2}{*}{$-$} & (0.1346) & (0.1056) & (0.1500) & (0.1201) & (0.0956) & (0.1307) & (0.1329) & (0.0969) & (0.1385)\\
\cmidrule{1-12}
 &  &  & 0.2688 & -0.0228 & -0.3656 & 0.2989 & -0.0135 & -0.3072 & 0.3133 & -0.0167 & -0.2207\\

 &  & \multirow{-2}{*}{idty} & (0.0563) & (0.0285) & (0.0406) & (0.0626) & (0.0316) & (0.0451) & (0.0688) & (0.0348) & (0.0496)\\

 &  &  & 0.5084 & -0.0233 & -0.3875 & 0.4928 & -0.0531 & -0.3629 & 0.4586 & -0.0922 & -0.3067\\

 & \multirow{-4}{*}{GEE} & \multirow{-2}{*}{logit} & (0.0707) & (0.0291) & (0.0378) & (0.0725) & (0.0277) & (0.0361) & (0.0736) & (0.0260) & (0.0340)\\

 &  &  & 0.0152 & -0.0017 & -0.1330 & 0.0172 & 0.0099 & -0.0488 & 0.0034 & 0.0092 & 0.0636\\

 &  & \multirow{-2}{*}{idty} & (0.0476) & (0.0373) & (0.0575) & (0.0529) & (0.0414) & (0.0639) & (0.0582) & (0.0455) & (0.0702)\\

 &  &  & 0.4094 & -0.0187 & -0.2920 & 0.3936 & -0.0487 & -0.2708 & 0.3606 & -0.0881 & -0.2193\\

 & \multirow{-4}{*}{CWGEE} & \multirow{-2}{*}{logit} & (0.0633) & (0.0260) & (0.0383) & (0.0677) & (0.0251) & (0.0393) & (0.0716) & (0.0239) & (0.0398)\\

 &  &  & 0.0031 & 0.0251 & -0.0282 & 0.0102 & 0.0176 & -0.0278 & 0.0030 & 0.0043 & -0.0074\\

\multirow{-10}{*}{\raggedleft\arraybackslash 200} & \multirow{-2}{*}{Lan et al.} & \multirow{-2}{*}{$-$} & (0.0616) & (0.0566) & (0.0777) & (0.0572) & (0.0506) & (0.0701) & (0.0635) & (0.0522) & (0.0712)\\
\bottomrule
\end{tabular}
}
\end{table}

\begin{table}
\caption{Bias and Monte Carlo (MC) error (i.e., MC standard deviation of the estimates, in parentheses) of the estimators of the conditional occupation probabilities for the simulation scenario where the cluster sizes are non-informative. Also, idty = identity link, logit = logit link.}
\label{tab: cond_est_nics}
\resizebox{1.0\textwidth}{!}{%
\begin{tabular}{rlllllllllll}
\toprule
\multicolumn{3}{c}{ } & \multicolumn{3}{c}{Healthy} & \multicolumn{3}{c}{Moderate} & \multicolumn{3}{c}{Severe} \\
\cmidrule(l{3pt}r{3pt}){4-6} \cmidrule(l{3pt}r{3pt}){7-9} \cmidrule(l{3pt}r{3pt}){10-12}
$m$ & Model & Link & $Z_2^{Q_1}$ & $Z_2^{Q_2}$ & $Z_2^{Q_3}$ & $Z_2^{Q_1}$ & $Z_2^{Q_2}$ & $Z_2^{Q_3}$ & $Z_2^{Q_1}$ & $Z_2^{Q_2}$ & $Z_2^{Q_3}$\\
\midrule
 &  &  & 0.0149 & -0.0127 & -0.1218 & -0.0011 & 0.0068 & -0.0275 & -0.0049 & 0.0181 & 0.0629\\

 &  & \multirow{-2}{*}{idty} & (0.0660) & (0.0415) & (0.0750) & (0.0733) & (0.0461) & (0.0833) & (0.0806) & (0.0507) & (0.0916)\\

 &  &  & -0.0780 & -0.0644 & -0.1018 & -0.1518 & -0.0835 & -0.0586 & -0.2116 & -0.1072 & -0.0189\\

 & \multirow{-4}{*}{GEE} & \multirow{-2}{*}{logit} & (0.1230) & (0.0266) & (0.1441) & (0.1315) & (0.0252) & (0.1569) & (0.1390) & (0.0236) & (0.1690)\\

 &  &  & 0.0144 & -0.0126 & -0.1213 & -0.0017 & 0.0069 & -0.0269 & -0.0056 & 0.0182 & 0.0635\\

 &  & \multirow{-2}{*}{idty} & (0.0653) & (0.0415) & (0.0742) & (0.0726) & (0.0461) & (0.0824) & (0.0798) & (0.0507) & (0.0907)\\

 &  &  & -0.0782 & -0.0643 & -0.1016 & -0.1520 & -0.0834 & -0.0583 & -0.2118 & -0.1071 & -0.0186\\

 & \multirow{-4}{*}{CWGEE} & \multirow{-2}{*}{logit} & (0.1224) & (0.0266) & (0.1435) & (0.1308) & (0.0253) & (0.1563) & (0.1383) & (0.0237) & (0.1683)\\

 &  &  & 0.0147 & 0.0220 & -0.0367 & -0.0009 & 0.0269 & -0.0261 & -0.0003 & 0.0233 & -0.0230\\

\multirow{-10}{*}{\raggedleft\arraybackslash 30} & \multirow{-2}{*}{Lan et al.} & \multirow{-2}{*}{$-$} & (0.0791) & (0.0506) & (0.0910) & (0.0778) & (0.0464) & (0.0863) & (0.0820) & (0.0480) & (0.0882)\\
\cmidrule{1-12}
 &  &  & 0.0142 & -0.0200 & -0.1141 & -0.0019 & -0.0013 & -0.0189 & -0.0058 & 0.0091 & 0.0723\\

 &  & \multirow{-2}{*}{idty} & (0.0282) & (0.0201) & (0.0334) & (0.0314) & (0.0223) & (0.0371) & (0.0345) & (0.0246) & (0.0408)\\

 &  &  & -0.1709 & -0.0835 & -0.1781 & -0.2484 & -0.1011 & -0.1415 & -0.3109 & -0.1232 & -0.1080\\

 & \multirow{-4}{*}{GEE} & \multirow{-2}{*}{logit} & (0.0438) & (0.0155) & (0.0999) & (0.0449) & (0.0140) & (0.1081) & (0.0455) & (0.0125) & (0.1157)\\

 &  &  & 0.0139 & -0.0200 & -0.1137 & -0.0023 & -0.0013 & -0.0184 & -0.0062 & 0.0091 & 0.0728\\

 &  & \multirow{-2}{*}{idty} & (0.0277) & (0.0203) & (0.0328) & (0.0307) & (0.0225) & (0.0364) & (0.0338) & (0.0248) & (0.0400)\\

 &  &  & -0.1711 & -0.0836 & -0.1779 & -0.2486 & -0.1011 & -0.1414 & -0.3110 & -0.1232 & -0.1079\\

 & \multirow{-4}{*}{CWGEE} & \multirow{-2}{*}{logit} & (0.0437) & (0.0155) & (0.0998) & (0.0448) & (0.0140) & (0.1080) & (0.0454) & (0.0125) & (0.1155)\\

 &  &  & 0.0055 & 0.0037 & -0.0092 & -0.0055 & 0.0094 & -0.0039 & -0.0015 & 0.0084 & -0.0069\\

\multirow{-10}{*}{\raggedleft\arraybackslash 200} & \multirow{-2}{*}{Lan et al.} & \multirow{-2}{*}{$-$} & (0.0360) & (0.0257) & (0.0420) & (0.0345) & (0.0236) & (0.0396) & (0.0352) & (0.0253) & (0.0404)\\
\bottomrule
\end{tabular}
}
\end{table}

For the simulation scenario where the cluster sizes are informative, Table \ref{tab: cond_est} shows the bias and Monte Carlo standard deviation of the estimates for each state in the three-state tracking model. The pseudo-value approach based on the CWGEE with an identity link performs slightly better than the inverse cluster size reweighted estimator of Lan et al. Both methods achieve the least bias among the competing methods and their corresponding bias decrease with an increasing number of clusters ($m=200$). As expected, transforming the pseudo-value responses to binary indicators does not improve the performance of the pseudo-value approach in estimating $\pi_\ell(t|z)$. Also, estimators from the pseudo-value approach formulated with estimating equations that were not reweighted by the inverse of the cluster size led to poor performance.

Table \ref{tab: cond_est_nics} shows the simulation results for the various estimators of $\pi_\ell(t|z)$ under the scenario where the cluster sizes are non-informative. The results show that reweighting the appropriate estimators by the inverse of the cluster size even when the cluster sizes are not informative still leads to valid estimation. The pseudo-value regression based on either CWGEE or the GEE with an identity link produces a similar performance to the method by Lan et al\cite{lan2017}. These estimators yield smaller bias and standard errors, which further decreases with an increasing number of clusters. Also, the pseudo-value regression with a logit link yields worse performance among the competing methods.

\subsection{Covariate inference based on the pseudo-value regression}
Suppose we are interested in quantifying the effect of more than one covariate on the SOP. For such an investigation, the method of Lan et al\cite{lan2017} will be inapplicable. In this subsection, we evaluate the performance of the pseudo-value approach for inference when multiple covariates are available. \textcolor{black}{We compute the pseudo-values at a fixed set of time points $(k=1,\hdots,10)$ equally spread across the event-time scale. For each state in the three-state tracking model, we initially conduct a power study for testing the hypothesis $\text{H}_0: \delta_1 = 0$ vs. $\text{H}_1: \delta_1 \neq 0$. The hypothesis test is based on fitting the pseudo-value regression model: $\mathbb{E}({Y}_{ij}(t_k)|\mathbf{Z}_{ij}) = \beta_{0,k} + \beta_1 Z_{1, i} + \beta_2 Z_{2,ij}$. The simulated power study is based on Wald-type tests where we vary $\delta_1$ in the interval [0, 0.75], and we set the nominal size $\alpha=0.05$. The test statistic for the Wald test is $z=\frac{\widehat{\beta}_1}{\sqrt{\widehat{\text{Var}}(\widehat{\beta}_1)}}$, where $\widehat{\text{Var}}(\widehat{\beta}_1)$ is the robust (sandwich) estimator of $\widehat{\beta}_1$. Note, $\beta_1$ is a non-linear function of $\delta_1$, such that $\text{H}_{0, \delta}: \delta_1 = 0$ and $\text{H}_{0, \beta}: \beta_1 = 0$ corresponds to the same null hypothesis of no covariate effect.} Again, we assess the relevance of reweighting the estimating equations (for fitting the regression models) by the inverse of the cluster size. Next, we also present results evaluating the performance of the estimating equations in estimating the pseudo-value regression coefficient $\beta_1$. \textcolor{black}{Our evaluations also include choosing three correlation structures, namely, independent, exchangeable, and AR1, in modeling the temporal correlations.}

\textcolor{black}{Tables \ref{tab3: ICS_tab} (and \ref{tab3: nonICS_tab}) present the empirical size and power estimates corresponding to the test $\text{H}_0: \delta_1 = 0$, when cluster sizes are informative (non-informative). From the estimated empirical sizes (in Table \ref{tab3: ICS_tab}), we infer that the GEE led to tests that were more biased than the CWGEE for each correlation structure. With increase in the number of clusters ($m=200$), the performance of GEE did not improve; however, the CWGEE led to tests that maintain the nominal size. Specifically, for the healthy and severe states, the CWGEE led to tests that were more powerful than the tests corresponding to the GEE. The power of the tests for the CWGEE models in these states approaches 1 as the number of clusters increases. However, under cluster non-informativeness (Table \ref{tab3: nonICS_tab}), the simulation results from both regression models are somewhat similar. Both models led to tests that maintain the nominal size with an increasing number of clusters, and the power of the tests approaches 1 with an increasing magnitude of the effect size. Furthermore, Tables \textcolor{red}{A.1} and \textcolor{red}{A.2} (in Web Appendix \textcolor{red}{A}) present the simulation results for the estimation of $\beta_1$. Compared to the CWGEE, we observe higher estimation bias and poorer coverage probabilities for the GEE models. The CWGEE yields coverage probabilities that are close to the nominal level, and estimated standard errors that are smaller than those of the GEE.}

\begin{table}[ht]
\caption{Empirical size (95\% confidence intervals), and power comparisons of hypothesis tests corresponding to the GEE and CWGEE models (under nominal $\alpha = 0.05$), for simulation scenario where the cluster sizes are informative. The pseudo-values were computed at 10 time points equally spread on the transition time scale. We consider three different correlation structures (independent, exchangeable, and AR1) for modeling the temporal correlation in the estimating equations.} 
\label{tab3: ICS_tab}
\centering
\resizebox{1.0\textwidth}{!}{%
\begin{tabular}[t]{rlllllll}
\toprule
\multicolumn{5}{c}{ } & \multicolumn{3}{c}{Power (under $\delta_1 > 0$)} \\
\cmidrule(l{3pt}r{3pt}){6-8}
$m$ & State & Cor. Structure & Model & Size (95\% CI) & 0.25 & 0.50 & 0.75\\
\midrule
 &  &  & GEE & 0.3270 (0.2979, 0.3561) & 0.1970 & 0.2690 & 0.4460\\

 &  & \multirow{-2}{*}{Independent} & CWGEE & 0.0640 (0.0488, 0.0792) & 0.1570 & 0.5040 & 0.8550\\

 &  &  & GEE & 0.3260 (0.2969, 0.3551) & 0.1940 & 0.2620 & 0.4430\\

 &  & \multirow{-2}{*}{Exchangeable} & CWGEE & 0.0650 (0.0497, 0.0803) & 0.1580 & 0.5040 & 0.8550\\

 &  &  & GEE & 0.3090 (0.2804, 0.3376) & 0.1840 & 0.2270 & 0.4120\\

 & \multirow{-6}{*}{Healthy} & \multirow{-2}{*}{AR1} & CWGEE & 0.0590 (0.0444, 0.0736) & 0.1550 & 0.4810 & 0.8040\\
\cmidrule{3-8}

 &  &  & GEE & 0.1980 (0.1733, 0.2227) & 0.1460 & 0.1270 & 0.1460\\

 &  & \multirow{-2}{*}{Independent} & CWGEE & 0.0710 (0.0551, 0.0869) & 0.0840 & 0.1090 & 0.0820\\

 &  &  & GEE & 0.1850 (0.1609, 0.2091) & 0.1440 & 0.1240 & 0.1420\\

 &  & \multirow{-2}{*}{Exchangeable} & CWGEE & 0.0720 (0.0560, 0.0880) & 0.0800 & 0.1030 & 0.0780\\

 &  &  & GEE & 0.1530 (0.1307, 0.1753) & 0.1260 & 0.1430 & 0.1710\\

 & \multirow{-6}{*}{Moderate} & \multirow{-2}{*}{AR1} & CWGEE & 0.0650 (0.0497, 0.0803) & 0.0880 & 0.0920 & 0.0860\\
\cmidrule{3-8}

 &  &  & GEE & 0.3310 (0.3018, 0.3602) & 0.1810 & 0.2230 & 0.3790\\

 &  & \multirow{-2}{*}{Independent} & CWGEE & 0.0810 (0.0641, 0.0979) & 0.1590 & 0.5010 & 0.8130\\

 &  &  & GEE & 0.3240 (0.2950, 0.3530) & 0.1790 & 0.2190 & 0.3760\\

 &  & \multirow{-2}{*}{Exchangeable} & CWGEE & 0.0760 (0.0596, 0.0924) & 0.1570 & 0.5020 & 0.8100\\

 &  &  & GEE & 0.3270 (0.2979, 0.3561) & 0.1700 & 0.2000 & 0.3650\\

\multirow{-18}{*}{\raggedleft\arraybackslash 30} & \multirow{-6}{*}{Severe} & \multirow{-2}{*}{AR1} & CWGEE & 0.0690 (0.0533, 0.0847) & 0.1450 & 0.5030 & 0.7690\\
\cmidrule{1-8}
 &  &  & GEE & 0.6350 (0.6052, 0.6648) & 0.2060 & 0.1520 & 0.4950\\

 &  & \multirow{-2}{*}{Independent} & CWGEE & 0.0480 (0.0348, 0.0612) & 0.6840 & 0.9880 & 0.9990\\

 &  &  & GEE & 0.6330 (0.6031, 0.6629) & 0.2050 & 0.1510 & 0.4960\\

 &  & \multirow{-2}{*}{Exchangeable} & CWGEE & 0.0480 (0.0348, 0.0612) & 0.6840 & 0.9880 & 0.9990\\

 &  &  & GEE & 0.6530 (0.6235, 0.6825) & 0.2090 & 0.1560 & 0.5010\\

 & \multirow{-6}{*}{Healthy} & \multirow{-2}{*}{AR1} & CWGEE & 0.0550 (0.0409, 0.0691) & 0.6340 & 0.9870 & 0.9980\\
\cmidrule{3-8}

 &  &  & GEE & 0.2820 (0.2541, 0.3099) & 0.1450 & 0.1110 & 0.1380\\

 &  & \multirow{-2}{*}{Independent} & CWGEE & 0.0460 (0.0330, 0.0590) & 0.1160 & 0.2380 & 0.2320\\

 &  &  & GEE & 0.2750 (0.2473, 0.3027) & 0.1430 & 0.1070 & 0.1380\\

 &  & \multirow{-2}{*}{Exchangeable} & CWGEE & 0.0450 (0.0322, 0.0578) & 0.1160 & 0.2330 & 0.2230\\

 &  &  & GEE & 0.2060 (0.1809, 0.2311) & 0.1200 & 0.0880 & 0.1300\\

 & \multirow{-6}{*}{Moderate} & \multirow{-2}{*}{AR1} & CWGEE & 0.0550 (0.0409, 0.0691) & 0.0840 & 0.1460 & 0.1230\\
\cmidrule{3-8}

 &  &  & GEE & 0.7600 (0.7335, 0.7865) & 0.2710 & 0.1140 & 0.4440\\

 &  & \multirow{-2}{*}{Independent} & CWGEE & 0.0600 (0.0453, 0.0747) & 0.7040 & 0.9890 & 0.9990\\

 &  &  & GEE & 0.7590 (0.7325, 0.7855) & 0.2700 & 0.1140 & 0.4420\\

 &  & \multirow{-2}{*}{Exchangeable} & CWGEE & 0.0600 (0.0453, 0.0747) & 0.7010 & 0.9890 & 0.9990\\

 &  &  & GEE & 0.7540 (0.7273, 0.7807) & 0.2700 & 0.1160 & 0.4600\\

\multirow{-18}{*}{\raggedleft\arraybackslash 200} & \multirow{-6}{*}{Severe} & \multirow{-2}{*}{AR1} & CWGEE & 0.0600 (0.0453, 0.0747) & 0.6700 & 0.9890 & 0.9980\\
\bottomrule
\end{tabular}
}
\end{table}

\FloatBarrier
\begin{table}[ht]
\caption{Empirical size (95\% confidence intervals), and power comparisons of hypothesis tests corresponding to the GEE and CWGEE models (nominal $\alpha = 0.05$) under the setting where the cluster sizes are non-informative. The pseudo-values were computed at 10 time points equally spread on the transition time scale. We consider three different correlation structures (independent, exchangeable, and AR1) for modeling the temporal correlation in the estimating equations.}
\label{tab3: nonICS_tab}
\centering
\resizebox{0.95\textwidth}{!}{%
\begin{tabular}[t]{rlllllll}
\toprule
\multicolumn{5}{c}{ } & \multicolumn{3}{c}{Power (under $\delta_1 > 0$)} \\
\cmidrule(l{3pt}r{3pt}){6-8}
$m$ & State & Cor. Structure & Model & Size (95\% CI) & 0.25 & 0.50 & 0.75\\
\midrule
 &  &  & GEE & 0.0700 (0.0542, 0.0858) & 0.2050 & 0.6590 & 0.9470\\

 &  & \multirow{-2}{*}{Independent} & CWGEE & 0.0620 (0.0471, 0.0769) & 0.1980 & 0.6650 & 0.9540\\

 &  &  & GEE & 0.0680 (0.0524, 0.0836) & 0.2080 & 0.6570 & 0.9460\\

 &  & \multirow{-2}{*}{Exchangeable} & CWGEE & 0.0620 (0.0471, 0.0769) & 0.1970 & 0.6630 & 0.9520\\

 &  &  & GEE & 0.0660 (0.0506, 0.0814) & 0.2190 & 0.6520 & 0.9430\\

 & \multirow{-6}{*}{Healthy} & \multirow{-2}{*}{AR1} & CWGEE & 0.0700 (0.0542, 0.0858) & 0.2220 & 0.6620 & 0.9510\\
\cmidrule{3-8}

 &  &  & GEE & 0.0660 (0.0506, 0.0814) & 0.1640 & 0.3210 & 0.3500\\

 &  & \multirow{-2}{*}{Independent} & CWGEE & 0.0640 (0.0488, 0.0792) & 0.1560 & 0.3230 & 0.3630\\

 &  &  & GEE & 0.0660 (0.0506, 0.0814) & 0.1620 & 0.3180 & 0.3440\\

 &  & \multirow{-2}{*}{Exchangeable} & CWGEE & 0.0640 (0.0488, 0.0792) & 0.1540 & 0.3210 & 0.3570\\

 &  &  & GEE & 0.0660 (0.0506, 0.0814) & 0.1350 & 0.2380 & 0.2690\\

 & \multirow{-6}{*}{Moderate} & \multirow{-2}{*}{AR1} & CWGEE & 0.0580 (0.0435, 0.0725) & 0.1350 & 0.2390 & 0.2720\\
\cmidrule{3-8}

 &  &  & GEE & 0.0700 (0.0542, 0.0858) & 0.2390 & 0.6800 & 0.9500\\

 &  & \multirow{-2}{*}{Independent} & CWGEE & 0.0650 (0.0497, 0.0803) & 0.2360 & 0.6970 & 0.9540\\

 &  &  & GEE & 0.0700 (0.0542, 0.0858) & 0.2390 & 0.6770 & 0.9480\\

 &  & \multirow{-2}{*}{Exchangeable} & CWGEE & 0.0650 (0.0497, 0.0803) & 0.2340 & 0.6940 & 0.9510\\

 &  &  & GEE & 0.0660 (0.0506, 0.0814) & 0.2320 & 0.6830 & 0.9470\\

\multirow{-18}{*}{\raggedleft\arraybackslash 30} & \multirow{-6}{*}{Severe} & \multirow{-2}{*}{AR1} & CWGEE & 0.0680 (0.0524, 0.0836) & 0.2340 & 0.6950 & 0.9430\\
\cmidrule{1-8}
 &  &  & GEE & 0.0610 (0.0462, 0.0758) & 0.8460 & 1.0000 & 1.0000\\

 &  & \multirow{-2}{*}{Independent} & CWGEE & 0.0630 (0.0479, 0.0781) & 0.8540 & 1.0000 & 1.0000\\

 &  &  & GEE & 0.0610 (0.0462, 0.0758) & 0.8460 & 1.0000 & 1.0000\\

 &  & \multirow{-2}{*}{Exchangeable} & CWGEE & 0.0620 (0.0471, 0.0769) & 0.8540 & 1.0000 & 1.0000\\

 &  &  & GEE & 0.0590 (0.0444, 0.0736) & 0.8460 & 1.0000 & 1.0000\\

 & \multirow{-6}{*}{Healthy} & \multirow{-2}{*}{AR1} & CWGEE & 0.0630 (0.0479, 0.0781) & 0.8510 & 1.0000 & 1.0000\\
\cmidrule{3-8}
 &  &  & GEE & 0.0470 (0.0339, 0.0601) & 0.5580 & 0.9450 & 0.9830\\

 &  & \multirow{-2}{*}{Independent} & CWGEE & 0.0560 (0.0417, 0.0703) & 0.5670 & 0.9460 & 0.9770\\

 &  &  & GEE & 0.0480 (0.0348, 0.0612) & 0.5550 & 0.9430 & 0.9820\\

 &  & \multirow{-2}{*}{Exchangeable} & CWGEE & 0.0550 (0.0409, 0.0691) & 0.5640 & 0.9430 & 0.9760\\

 &  &  & GEE & 0.0490 (0.0356, 0.0624) & 0.5020 & 0.8950 & 0.9540\\

 & \multirow{-6}{*}{Moderate} & \multirow{-2}{*}{AR1} & CWGEE & 0.0580 (0.0435, 0.0725) & 0.4930 & 0.8920 & 0.9480\\
\cmidrule{3-8}

 &  &  & GEE & 0.0550 (0.0409, 0.0691) & 0.8840 & 1.0000 & 1.0000\\

 &  & \multirow{-2}{*}{Independent} & CWGEE & 0.0610 (0.0462, 0.0758) & 0.8860 & 1.0000 & 1.0000\\

 &  &  & GEE & 0.0520 (0.0382, 0.0658) & 0.8840 & 1.0000 & 1.0000\\

 &  & \multirow{-2}{*}{Exchangeable} & CWGEE & 0.0630 (0.0479, 0.0781) & 0.8850 & 1.0000 & 1.0000\\

 &  &  & GEE & 0.0570 (0.0426, 0.0714) & 0.8810 & 1.0000 & 1.0000\\

\multirow{-18}{*}{\raggedleft\arraybackslash 200} & \multirow{-6}{*}{Severe} & \multirow{-2}{*}{AR1} & CWGEE & 0.0590 (0.0444, 0.0736) & 0.8810 & 1.0000 & 1.0000\\
\bottomrule
\end{tabular}
}
\end{table}
\FloatBarrier

\textcolor{black}{Note that the responses of the estimating equations are the jackknife pseudo-values obtained from unweighted marginal estimators of the SOP. Under the scenario where the cluster sizes are either informative or non-informative, the simulation results for the pseudo-value regression based on current-status data show that reweighting the estimating equation by the inverse of the cluster size led to reliable inferential results. Also, the simulation results corresponding to each estimating equation do not vary significantly across the different correlation structures used in this setting.}

\textcolor{black}{Motivated by a reviewer's comment, Table \textcolor{red}{B.1} in Web-supplement B summarizes our findings from a sensitivity assessment to the choice of the number of time points (at which the pseudo-values are calculated), spread equally across the transition scale using synthetic data. The data simulation scheme remains the same, as described earlier. We observe that the empirical coverage probabilities are fairly decent for small number of clusters ($m = 30$), with marked improvement when the number of clusters increases ($m = 200$). Similar findings (i.e., performance improvement with increasing $m$) are observed while evaluating the other metrics, such as the Monte Carlo standard deviation, average estimated standard error, mean squared error, and bias. This is in line with the underlying large sample theory. However, for a fixed $m$, there does not seem to be any significant improvement in estimation precision for calculating the pseudo-values using more time points beyond seven to ten. Henceforth, in subsequent data application in Section \ref{sec: application}, we restricted our attention to 10 time-points.}

\textcolor{black}{In Web-supplement \textcolor{red}{C}, we also present simulation results for the scenario where there are known groups within the clusters, and the intra-cluster group sizes are informative. For details on data generation, refer to Web-supplement \textcolor{red}{C}. The empirical estimates of the size (with 95\% confidence intervals) and power are shown in Table \textcolor{red}{C.1}. In terms of the effect size, the CWGEE\cite{williamson2003} and GEE\cite{liang1986} models led to tests that are biased, even for large number of clusters ($m=200$). However, the DWGEE\citep{huang2011} led to tests that maintain the nominal size and are more powerful than the other models. Table \textcolor{red}{C.2} presents simulation results for estimating $\beta_1$. The GEE and CWGEE models yield poor coverage probabilities and higher estimation bias, whereas the DWGEE is approximately unbiased and yields coverage probabilities that are close to the nominal level. Also, the estimated standard errors from the DWGEE fit tend to be smaller than those of the GEE and CWGEE. However, the results corresponding to each correlation structure are nearly identical. The results reveal that one will obtain reliable inference by reweighting the contribution of a cluster unit to the estimating equations via the inverse of the number of observations within the cluster having the same group membership as the cluster unit.}


\section{Application: GAAD Data}\label{sec: application}
We illustrate the implementation of the proposed approach in analyzing the data set from the GAAD study introduced in Section \ref{sec: intro}. The GAAD study aimed to explore the relationship between PD and diabetes (determined by the marker glycosylated hemoglobin or HbA1c) among the Gullah-speaking African-Americans living on the sea islands of South Carolina \cite{fernandes2009}. PD incidence was determined by calculating the clinical attachment level (CAL) at six sites of each tooth (excluding the third molars) in a patient's mouth using a periodontal probe. CAL is defined as the distance between the cementoenamel junction and the base of the tooth's pocket. Following the guidelines from the American Association of Periodontology \cite{armitage1999}, the four respective states in the multistate tracking model that describes the severity of PD are as follows: CAL = 0 denotes healthy, $\text{CAL} = \{1, 2\}$ denotes slight PD, $\text{CAL} = \{3, 4\}$ denotes moderate PD, \textcolor{black}{and CAL $\geq 5$ denotes severe PD}.

In this cross-sectional study, although the time of visit to the dental office was the same for all patients, the exact examination time will not be the same due to the variation in permanent dentition times of different tooth types. We compute the CS time for a tooth site as the difference between the patient's age and the approximate permanent dentition times of US adults, obtained from eruption charts maintained by the American Dental Association; see link \url{https://www.mouthhealthy.org/en/az-topics/e/eruption-charts}. 

The study investigators were interested in quantifying the association of risk factors to the latent time to PD incidence at the tooth-site level, and, in particular, the evaluation of HbA1c with PD incidence -- a well-explored topic in PD research \cite{chapple2013diabetes}. The data comprise $m=288$ patients (clusters) with at least one tooth present at the time of examination. From the marginal (no covariates) analysis presented in Section \ref{sec: intro}, patients with fewer available teeth-sites were at more risk of PD than patients with more teeth-sites. This finding indicates the potential presence of ICS in this data set. Each patient had at least one tooth site that was identified with some form of PD incidence. We modeled the probability that a typical tooth site in a typical patient is healthy as a function of five risk factors namely: jaw location/maxilla (0=lower jaw, 1=upper jaw), smoking status (0=non-smoker, 1=smoker), gender (0=male, 1=female), HbA1c (0=controlled, 1=high level) and body mass index (BMI). Among the 288 participants, 76\% were female, 31\% were current or previous smokers, and 59\% had high HbA1c levels. To implement the pseudo-value regression, we compute the pseudo-values at $10$ time points. The pseudo-values were calculated using the unweighted marginal estimator of the occupation probability of occupying the healthy state. The time points were equally spaced across the observed CS times; this gives us an ideal summary of the distribution of the observed event times. We fit both GEE and CWGEE models with an identity link function so that the estimated covariate effects are interpreted in terms of absolute risk reduction (ARR) \cite{ambrogi2008}. Further, we use an AR-1 correlation structure to describe the temporal correlation induced by computing the pseudo-values at multiple time points, while the correlation parameter is estimated by the two-stage quasi-least squares approach \cite{chaganty1997}. 

\textcolor{black}{In Table \ref{tab3: PDstudy}, we report the estimates of the regression coefficients, sandwich standard errors, and corresponding $p$-values for modeling tooth sites occupying healthy and severe states, respectively.} From the CWGEE model, the HbA1c marker is significantly associated (at the $\alpha=0.05$ significance level) with the probability of occupying the healthy tooth state. Specifically, when compared to a typical tooth site from a patient with well-controlled HbA1c, a typical tooth site from a patient with high/uncontrolled HbA1c (adjusted for gender, smoking status, BMI, and maxilla-located) is on average, 10\% less likely to be healthy. Note that the estimate of HbA1c, though negative, was not significant for the usual GEE model. Findings from the Lam et al \cite{lam2021} study, which analyzed the traditional current status PD responses (not multistate) recently from the GAAD data using a class of semiparametric transformation cure models, also revealed that HbA1c is a significant predictor associated with healthy teeth. With the exception of gender (though, not significant), the estimated effects from both GEE and CWGEE models were similar in sign, but with varying magnitudes. However, under our GEE model, none of the available covariates were significantly associated with the probability of a tooth site staying healthy. \textcolor{black}{On the other hand, both regression models do not identify any risk factor to be significantly associated with occupying the severe state.}

\FloatBarrier
\begin{table}[ht]
\caption{Parameter estimates, standard error (SE), and p-values from the CWGEE and GEE pseudo-value regression models for fitting the probability that a tooth site occupies a defined state (`Healthy', or `Severe') from the GAAD study.}
\label{tab3: PDstudy}
\centering
\begin{tabular}[t]{llrrrrrr}
\toprule
\multicolumn{2}{c}{ } & \multicolumn{3}{c}{CWGEE} & \multicolumn{3}{c}{GEE} \\
\cmidrule(l{3pt}r{3pt}){3-5} \cmidrule(l{3pt}r{3pt}){6-8}
State & Covariates & Estimate & SE & $p$-value & Estimate & SE & $p$-value\\
\midrule
 & BMI & -0.0012 & 0.0021 & 0.5683 & -0.0022 & 0.0023 & 0.3295\\

 & Female & -0.0324 & 0.0489 & 0.5073 & -0.0170 & 0.0476 & 0.7214\\

 & HbA1c & -0.1089 & 0.0520 & 0.0362 & -0.0986 & 0.0636 & 0.1212\\

 & Maxilla & -0.0510 & 0.0464 & 0.2720 & -0.0633 & 0.0498 & 0.2032\\

\multirow{-5}{*}{Healthy} & Smoker & -0.0190 & 0.0491 & 0.6988 & -0.0521 & 0.0587 & 0.3746\\
\cmidrule{1-8}
 & BMI & -0.0036 & 0.0027 & 0.1770 & 0.0001 & 0.0013 & 0.9566\\

 & Female & -0.0176 & 0.0599 & 0.7693 & -0.0174 & 0.0240 & 0.4676\\

 & HbA1c & 0.0767 & 0.0472 & 0.1042 & 0.0426 & 0.0228 & 0.0618\\

 & Maxilla & 0.0011 & 0.0438 & 0.9793 & 0.0189 & 0.0167 & 0.2592\\

\multirow{-5}{*}{Severe} & Smoker & -0.0126 & 0.0628 & 0.8415 & -0.0022 & 0.0232 & 0.9249\\
\bottomrule
\end{tabular}
\end{table}
\FloatBarrier

\section{Discussion}\label{sec: discussion}
Motivated by a PD study, this paper carefully extends the pseudo-value approach to conduct a cross-sectional risk assessment for cluster-correlated CS data in a multistate setting. The presence of severe censoring induced by a single observation of the study units at a random inspection time, clustering of the study units within defined clusters, and the potential association of the cluster sizes with the transition outcomes further complicate the analysis of the multistate data. We tackle this via jackknife pseudo-values, which initially compute the pseudo-values that are based on (unweighted) marginal estimators of the SOP. Next, we fit the pseudo-values to the available covariates using both GEE and CWGEE models. In the case where the cluster sizes are non-informative, both regression models are valid. However, the GEE model yields biased estimates when the cluster sizes are informative.  The proposed approach can be generalized to other epidemiological settings that generate cluster-correlated CS data. 

\textcolor{black}{Alternatively, one may consider using the within-cluster-resampling \cite{hoffman2001} approach. For this approach, we create a dataset by randomly selecting a cluster unit from each of the $m$ clusters; the pseudo-values are then calculated based on the unweighted marginal estimates of the SOP for this dataset. The traditional GEE is then used to perform the regression analysis of the pseudo-values on the covariates in the resampled data. This procedure is repeated, say $Q$ number of times, and the estimated regression parameters are obtained by averaging over the $Q$ estimates from the replicate data sets. Our proposed approach (CWGEE with pseudo-value responses that are based on unweighted marginal estimates) are expected to be similar to the WCR approach. However, we found the WCR method to be computationally intensive, with the final estimates to be unstable, which was also pointed out earlier in the literature \cite{williamson2003}.}

From our current investigation, there are several future directions to consider. First, we have described our modeling approach based on the SOP; however, the techniques proposed in this article can be easily extended to other temporal parameters of interest in either competing risks or the MSM framework. For instance, one can easily estimate the covariate effects on the transition probabilities by using the pseudo-values that are based on the estimator given in (\ref{eqn2: transProb}). Other temporal parameters that may be of interest in the multistate model framework include the restricted mean survival time and the absolute risk. In the competing risks framework, one can also extend the method to model covariate effects on the cumulative incidence function. Next, our proposal focused on case-I interval censoring, however, it may be worthwhile to consider scenarios where the clustered event times are subject to more general (such as case-II) interval censoring. Furthermore, we assumed that the interval censoring is independent of the covariates. Under violations of that assumption, one may consider extending our work to the dependent censoring scenarios, where the clustered-correlated inspection times may depend on either covariates or the states of the multistate model. 

\textcolor{black}{In settings with right-censored data, several papers \cite{jacobsen2016, overgaard2017, overgaard2018} have noted that the sandwich variance estimator (as in (\ref{eqn: sandwich})) for the pseudo-value inference may be conservative, yet, the resulting bias may have an insignificant impact \citep{jacobsen2016}, in practice. For the CS setting where the marginal estimators are constructed via nonparametric smoothing, it may be worthwhile to conduct a theoretical study of the (sandwich) variance estimators, and whether they yield conservative inference as in the case of right-censored data. This is also part of our future work, to be explored elsewhere.}

\bibliography{arXiv.bib}

\begin{thebibliography}{3}
\providecommand{\natexlab}[1]{#1}
\providecommand{\url}[1]{\texttt{#1}}
\expandafter\ifx\csname urlstyle\endcsname\relax
  \providecommand{\doi}[1]{doi: #1}\else
  \providecommand{\doi}{doi: \begingroup \urlstyle{rm}\Url}\fi

\bibitem[Liang and Zeger(1986)]{liang1986b}
Kung-Yee Liang and Scott~L Zeger.
\newblock Longitudinal data analysis using generalized linear models.
\newblock \emph{Biometrika}, 73\penalty0 (1):\penalty0 13--22, 1986.

\bibitem[Williamson et~al.(2003)Williamson, Datta, and Satten]{williamson2003b}
John~M Williamson, Somnath Datta, and Glen~A Satten.
\newblock Marginal analyses of clustered data when cluster size is informative.
\newblock \emph{Biometrics}, 59\penalty0 (1):\penalty0 36--42, 2003.

\bibitem[Huang and Leroux(2011)]{huang2011b}
Ying Huang and Brian Leroux.
\newblock Informative cluster sizes for subcluster-level covariates and
  weighted generalized estimating equations.
\newblock \emph{Biometrics}, 67\penalty0 (3):\penalty0 843--851, 2011.

\end{thebibliography}


\begin{thebibliography}{37}
\providecommand{\natexlab}[1]{#1}
\providecommand{\url}[1]{\texttt{#1}}
\expandafter\ifx\csname urlstyle\endcsname\relax
  \providecommand{\doi}[1]{doi: #1}\else
  \providecommand{\doi}{doi: \begingroup \urlstyle{rm}\Url}\fi

\bibitem[Andersen and Keiding(2002)]{andersen2002multi}
Per~Kragh Andersen and Niels Keiding.
\newblock Multi-state models for event history analysis.
\newblock \emph{Statistical Methods in Medical Research}, 11\penalty0
  (2):\penalty0 91--115, 2002.

\bibitem[Mitani et~al.(2021)Mitani, Kaye, and Nelson]{mitani2021marginal}
AA~Mitani, EK~Kaye, and KP~Nelson.
\newblock Marginal analysis of multiple outcomes with informative cluster size.
\newblock \emph{Biometrics}, 77\penalty0 (1):\penalty0 271--282, 2021.

\bibitem[Lan et~al.(2017)Lan, Bandyopadhyay, and Datta]{lan2017}
Ling Lan, Dipankar Bandyopadhyay, and Somnath Datta.
\newblock Non-parametric regression in clustered multistate current status data
  with informative cluster size.
\newblock \emph{Statistica Neerlandica}, 71\penalty0 (1):\penalty0 31--57,
  2017.

\bibitem[Fernandes et~al.(2009)Fernandes, Wiegand, Salinas, Grossi, Sanders,
  Lopes-Virella, and Slate]{fernandes2009}
Jyotika~K Fernandes, Ryan~E Wiegand, Carlos~F Salinas, Sara~G Grossi, John~J
  Sanders, Maria~F Lopes-Virella, and Elizabeth~H Slate.
\newblock {Periodontal disease status in Gullah African Americans with type 2
  diabetes living in South Carolina}.
\newblock \emph{Journal of Periodontology}, 80\penalty0 (7):\penalty0
  1062--1068, 2009.

\bibitem[Mdala et~al.(2014)Mdala, Olsen, Haffajee, Socransky, Thoresen, and
  de~Blasio]{mdala2014comparing}
Ibrahimu Mdala, Ingar Olsen, Anne~D Haffajee, Sigmund~S Socransky, Magne
  Thoresen, and Birgitte~Freiesleben de~Blasio.
\newblock {Comparing CAL and PD for predicting periodontal disease progression
  in healthy sites of patients with chronic periodontitis using multi-state
  Markov models}.
\newblock \emph{Journal of Clinical Periodontology}, 41\penalty0 (9):\penalty0
  837, 2014.

\bibitem[Jewell and van~der Laan(2003)]{jewell2003}
Nicholas~P Jewell and Mark van~der Laan.
\newblock Current status data: Review, recent developments and open problems.
\newblock \emph{Handbook of statistics}, 23:\penalty0 625--642, 2003.

\bibitem[Shiboski(1998)]{shiboski1998}
Stephen~C Shiboski.
\newblock Generalized additive models for current status data.
\newblock \emph{Lifetime data analysis}, 4\penalty0 (1):\penalty0 29--50, 1998.

\bibitem[Sun(2006)]{sun2006}
Jianguo Sun.
\newblock \emph{The Statistical Analysis of Interval-Censored Failure Time
  Data}.
\newblock Springer, New York, 2006.

\bibitem[Chen et~al.(2009)Chen, Tong, and Sun]{chen2009}
Man-Hua Chen, Xingwei Tong, and Jianguo Sun.
\newblock A frailty model approach for regression analysis of multivariate
  current status data.
\newblock \emph{Statistics in medicine}, 28\penalty0 (27):\penalty0 3424--3436,
  2009.

\bibitem[Li et~al.(2022)Li, Ma, Sun, and Tang]{li2022new}
Huiqiong Li, Chenchen Ma, Jianguo Sun, and Niansheng Tang.
\newblock A new approach for regression analysis of multivariate current status
  data with informative censoring.
\newblock \emph{Communications in Mathematics and Statistics}, 2022.
\newblock in press.

\bibitem[Andersen et~al.(2003)Andersen, Klein, and Rosth{\o}j]{andersen2003}
Per~Kragh Andersen, John~P Klein, and Susanne Rosth{\o}j.
\newblock Generalised linear models for correlated pseudo-observations, with
  applications to multi-state models.
\newblock \emph{Biometrika}, 90\penalty0 (1):\penalty0 15--27, 2003.

\bibitem[Datta and Sundaram(2006)]{datta2006}
Somnath Datta and Rajeshwari Sundaram.
\newblock Nonparametric estimation of stage occupation probabilities in a
  multistage model with current status data.
\newblock \emph{Biometrics}, 62\penalty0 (3):\penalty0 829--837, 2006.

\bibitem[Anyaso-Samuel and Datta(2022)]{anyasosamuel2022}
Samuel Anyaso-Samuel and Somnath Datta.
\newblock Adjusting for informative cluster size in pseudo-value based
  regression approaches with clustered time to event data, 2022.
\newblock URL \url{https://arxiv.org/abs/2210.13410}.

\bibitem[Barlow et~al.(1972)Barlow, Bartholomew, Bremner, and
  Brunk]{barlow1972}
R.E. Barlow, D.J. Bartholomew, J.M. Bremner, and H.D. Brunk.
\newblock \emph{Statistical Inference Under Order Restrictions: The Theory and
  Application of Isotonic Regression}.
\newblock J. Wiley, 1972.

\bibitem[Mukerjee(1988)]{mukerjee1988}
Hari Mukerjee.
\newblock Monotone nonparametric regression.
\newblock \emph{The Annals of Statistics}, pages 741--750, 1988.

\bibitem[Nadaraya(1964)]{nadaraya1964}
Elizbar~A Nadaraya.
\newblock On estimating regression.
\newblock \emph{Theory of Probability \& Its Applications}, 9\penalty0
  (1):\penalty0 141--142, 1964.

\bibitem[Watson(1964)]{watson1964}
Geoffrey~S Watson.
\newblock Smooth regression analysis.
\newblock \emph{Sankhy{\=a}: The Indian Journal of Statistics, Series A}, pages
  359--372, 1964.

\bibitem[Wand and Jones(1994)]{wand1994}
Matt~P Wand and M~Chris Jones.
\newblock \emph{Kernel smoothing}.
\newblock CRC press, 1994.

\bibitem[Andersen et~al.(1993)Andersen, Borgan, Gill, and
  Keiding]{andersen1993}
Per~K Andersen, Ornulf Borgan, Richard~D Gill, and Niels Keiding.
\newblock \emph{Statistical models based on counting processes}.
\newblock Springer Science \& Business Media, 1993.

\bibitem[Datta and Satten(2001)]{datta2001}
Somnath Datta and Glen~A Satten.
\newblock {Validity of the Aalen--Johansen estimators of stage occupation
  probabilities and Nelson--Aalen estimators of integrated transition hazards
  for non-Markov models}.
\newblock \emph{Statistics \& probability letters}, 55\penalty0 (4):\penalty0
  403--411, 2001.

\bibitem[Klein and Andersen(2005)]{klein2005}
John~P Klein and Per~Kragh Andersen.
\newblock Regression modeling of competing risks data based on pseudovalues of
  the cumulative incidence function.
\newblock \emph{Biometrics}, 61\penalty0 (1):\penalty0 223--229, 2005.

\bibitem[Andersen and Pohar~Perme(2010)]{andersen2010}
Per~Kragh Andersen and Maja Pohar~Perme.
\newblock Pseudo-observations in survival analysis.
\newblock \emph{Statistical methods in medical research}, 19\penalty0
  (1):\penalty0 71--99, 2010.

\bibitem[Wang et~al.(2011)Wang, Kong, and Datta]{wang2011}
Ming Wang, Maiying Kong, and Somnath Datta.
\newblock Inference for marginal linear models for clustered longitudinal data
  with potentially informative cluster sizes.
\newblock \emph{Statistical Methods in Medical Research}, 20\penalty0
  (4):\penalty0 347--367, 2011.

\bibitem[Liang and Zeger(1986)]{liang1986}
Kung-Yee Liang and Scott~L Zeger.
\newblock Longitudinal data analysis using generalized linear models.
\newblock \emph{Biometrika}, 73\penalty0 (1):\penalty0 13--22, 1986.

\bibitem[Williamson et~al.(2003)Williamson, Datta, and Satten]{williamson2003}
John~M Williamson, Somnath Datta, and Glen~A Satten.
\newblock Marginal analyses of clustered data when cluster size is informative.
\newblock \emph{Biometrics}, 59\penalty0 (1):\penalty0 36--42, 2003.

\bibitem[Dutta and Datta(2016)]{dutta2016}
Sandipan Dutta and Somnath Datta.
\newblock A rank-sum test for clustered data when the number of subjects in a
  group within a cluster is informative.
\newblock \emph{Biometrics}, 72\penalty0 (2):\penalty0 432--440, 2016.

\bibitem[Huang and Leroux(2011)]{huang2011}
Ying Huang and Brian Leroux.
\newblock Informative cluster sizes for subcluster-level covariates and
  weighted generalized estimating equations.
\newblock \emph{Biometrics}, 67\penalty0 (3):\penalty0 843--851, 2011.

\bibitem[Armitage(1999)]{armitage1999}
Gary~C Armitage.
\newblock Development of a classification system for periodontal diseases and
  conditions.
\newblock \emph{Annals of periodontology}, 4\penalty0 (1):\penalty0 1--6, 1999.

\bibitem[Chapple et~al.(2013)Chapple, Genco, and working group 2 of the~joint
  EFP/AAP~workshop]{chapple2013diabetes}
Iain~LC Chapple, Robert Genco, and working group 2 of the~joint
  EFP/AAP~workshop.
\newblock {Diabetes and Periodontal Diseases: Consensus Report of the Joint
  EFP/AAP Workshop on Periodontitis and Systemic Diseases}.
\newblock \emph{Journal of Periodontology}, 84:\penalty0 S106--S112, 2013.

\bibitem[Ambrogi et~al.(2008)Ambrogi, Biganzoli, and Boracchi]{ambrogi2008}
Federico Ambrogi, Elia Biganzoli, and Patrizia Boracchi.
\newblock Estimates of clinically useful measures in competing risks survival
  analysis.
\newblock \emph{Statistics in medicine}, 27\penalty0 (30):\penalty0 6407--6425,
  2008.

\bibitem[Chaganty(1997)]{chaganty1997}
N~Rao Chaganty.
\newblock An alternative approach to the analysis of longitudinal data via
  generalized estimating equations.
\newblock \emph{Journal of Statistical Planning and Inference}, 63\penalty0
  (1):\penalty0 39--54, 1997.

\bibitem[Lam et~al.(2021)Lam, Lee, Wong, and Bandyopadhyay]{lam2021}
Kwok~Fai Lam, Chun~Yin Lee, Kin~Yau Wong, and Dipankar Bandyopadhyay.
\newblock Marginal analysis of current status data with informative cluster
  size using a class of semiparametric transformation cure models.
\newblock \emph{Statistics in Medicine}, 40\penalty0 (10):\penalty0 2400--2412,
  2021.

\bibitem[Hoffman et~al.(2001)Hoffman, Sen, and Weinberg]{hoffman2001}
Elaine~B Hoffman, Pranab~K Sen, and Clarice~R Weinberg.
\newblock Within-cluster resampling.
\newblock \emph{Biometrika}, 88\penalty0 (4):\penalty0 1121--1134, 2001.

\bibitem[Jacobsen and Martinussen(2016)]{jacobsen2016}
Martin Jacobsen and Torben Martinussen.
\newblock A note on the large sample properties of estimators based on
  generalized linear models for correlated pseudo-observations.
\newblock \emph{Scandinavian Journal of Statistics}, 43\penalty0 (3):\penalty0
  845--862, 2016.

\bibitem[Overgaard et~al.(2017)Overgaard, Parner, Pedersen,
  et~al.]{overgaard2017}
Morten Overgaard, Erik~Thorlund Parner, Jan Pedersen, et~al.
\newblock Asymptotic theory of generalized estimating equations based on
  jack-knife pseudo-observations.
\newblock \emph{Annals of Statistics}, 45\penalty0 (5):\penalty0 1988--2015,
  2017.

\bibitem[Overgaard et~al.(2018)Overgaard, Parner, and Pedersen]{overgaard2018}
Morten Overgaard, Erik~Thorlund Parner, and Jan Pedersen.
\newblock Estimating the variance in a pseudo-observation scheme with competing
  risks.
\newblock \emph{Scandinavian Journal of Statistics}, 45\penalty0 (4):\penalty0
  923--940, 2018.

\bibitem[Nevalainen et~al.(2014)Nevalainen, Datta, and Oja]{nevalainen2014}
Jaakko Nevalainen, Somnath Datta, and Hannu Oja.
\newblock Inference on the marginal distribution of clustered data with
  informative cluster size.
\newblock \emph{Statistical Papers}, 55\penalty0 (1):\penalty0 71--92, 2014.

\end{thebibliography}

\vspace{-0.5cm}
\begin{appendices}
\section{Inverse cluster size marginalization for the pseudo-value approach} \label{app: appenA}
Consider the case where the patients indexed by $j$ are clustered within distinct clusters indexed by $i$, and the cluster sizes $n_i$ are potentially informative. For a marginal analysis, inverse cluster size reweighting has been suggested in the literature \cite{nevalainen2014} to correct for the bias induced by the informative cluster sizes. Define the independent and identically distributed (IID) random elements $\mathbb{V}_{i} = \{n_i, X_{i1},\ldots, X_{in_i},\mathbf{Z}_{i1},\ldots,\mathbf{Z}_{in_i}\},\ i=1,\ldots,m$, and assume that $(X_{ij},\ \mathbf{Z}_{ij})$ within a given cluster $i$ are exchangeable given the cluster size $n_i$. Let $Y_{IJ}(t)$ and $Z_{IJ}$ respectively denote the pseudo-value and the corresponding continuous covariate for patient $J$ in cluster $I$, where $I$ is a randomly selected cluster from the sample with $m$ clusters, and $J$ is sampled from $\{1,...,n_I\}$ with probability $1/n_I$. Now, we show that by the inverse cluster size marginalization over the clusters, $\mathbb{E}\{Y_{IJ}(t) \ |\ Z_{IJ} = z\}$ will eventually lead to the SOP conditioned on the given covariate. Using the expression of the pseudo-value as shown in (\ref{eqn: pseudo1}) and conditioning on $I=i$ and $J=j$, we have
\begin{align*} 
    \mathbb{E}\{Y_{IJ}(t) \ |\ Z_{IJ} = z\}  &= \mathbb{E} \Big[\frac{1}{m} \sum^{m}_{i=1} \frac{1}{n_i} \sum^{n_i}_{j=1} \mathbb{E} \Big( \frac{1}{2h} \mathcal{I}\{X_{ij}(C_{ij}) = \ell,\ t-h < C_{ij} < t+h\} \ | \ Z_{ij}=z \Big)\Big]\\
    &\approx \mathbb{E} \Big[\frac{1}{m} \sum^{m}_{i=1} \frac{1}{n_i} \sum^{n_i}_{j=1} \mathbb{E} \Big(\mathcal{I}\{X_{ij}(t) = \ell\} \ | \ Z_{ij}=z \Big)\Big]\\
    &= \mathbb{E} \Big(\mathcal{I}\{X_{IJ}(t) = \ell\} \ | \ Z_{IJ}=z \Big)\\
    &= \mathbb{P} \Big(X_{IJ}(t) = \ell \ | \ Z_{IJ}=z \Big),
\end{align*}
where $\approx$ means equality up to $o(1)$ terms. The second approximation holds because the bias term in kernel smoothing is $o(1)$.

The inverse cluster size marginalization described above is appropriate for inference when the cluster sizes are informative. Suppose, we have two groups in each cluster ($G \in \{0,1\}$), and the intra-cluster group (ICG) sizes are informative. Let the index $J$ in $(Y_{IJ}, Z_{IJ})$ denote that a patient is sampled from cluster $I$ with probability $1/(2n_{I0})$ for those with $G=0$, and with probability $1/(2n_{I1})$ for those with $G=1$. The results shown above can be extended to the case with informative ICG sizes where the marginalization over the clusters is performed by reweighting each observation by $1/(2n_{IG_{IJ}})$.
\end{appendices}

\end{document}


\maketitle

\section{Estimation of $\beta_{1}$} \label{websec:beta1}

\begin{table}[ht]
\centering
\tiny{
\caption{\small{Simulation results for the pseudo-value regression parameter $\beta_1$ under the setting where the cluster sizes are informative. The pseudo-values were computed at 10 time points equally spread on the transition time scale. We consider three different correlation structures (independent, exchangeable, and AR1) for modeling the temporal correlation in the estimating equations.}} \label{web:bias_ICSm30_200} \vspace{-0.1in} 
\resizebox{1.0\textwidth}{!}{%
\begin{tabular}[t]{lrllrllll}
\toprule
$m$ & State & Cor. Structure & Model & Bias & MCSD & ASE & MSE & Coverage\\
\midrule
 &  &  & GEE & -13.2235 & 20.7509 & 13.7326 & 6.0503 & 0.7270\\

 &  & \multirow[t]{-2}{*}{\raggedright\arraybackslash Independent} & CWGEE & -1.3316 & 10.7906 & 9.7749 & 1.1809 & 0.9310\\

 &  &  & GEE & -13.2418 & 20.6278 & 13.7375 & 6.0043 & 0.7250\\

 &  & \multirow[t]{-2}{*}{\raggedright\arraybackslash Exchangeable} & CWGEE & -1.3480 & 10.7765 & 9.7751 & 1.1783 & 0.9310\\

 &  &  & GEE & -13.5013 & 18.9539 & 12.8423 & 5.4118 & 0.7140\\

 & \multirow[t]{-6}{*}{\raggedright\arraybackslash Healthy} & \multirow[t]{-2}{*}{\raggedright\arraybackslash AR1} & CWGEE & -1.5488 & 11.4314 & 10.3031 & 1.3295 & 0.9360\\

 &  &  & GEE & -5.1133 & 8.3074 & 7.0390 & 0.9509 & 0.7820\\

 &  & \multirow[t]{-2}{*}{\raggedright\arraybackslash Independent} & CWGEE & -1.1856 & 7.9035 & 7.1393 & 0.6381 & 0.8830\\

 &  &  & GEE & -5.0343 & 8.2425 & 7.0413 & 0.9321 & 0.7860\\

 &  & \multirow[t]{-2}{*}{\raggedright\arraybackslash Exchangeable} & CWGEE & -1.2436 & 7.8584 & 7.1395 & 0.6324 & 0.8840\\

 &  &  & GEE & -3.9298 & 8.1505 & 6.8458 & 0.8181 & 0.8020\\

 & \multirow[t]{-6}{*}{\raggedright\arraybackslash Moderate} & \multirow[t]{-2}{*}{\raggedright\arraybackslash AR1} & CWGEE & -2.0008 & 7.9968 & 7.1281 & 0.6789 & 0.8930\\

 &  &  & GEE & 18.3367 & 21.0692 & 15.4045 & 7.7970 & 0.6810\\

 &  & \multirow[t]{-2}{*}{\raggedright\arraybackslash Independent} & CWGEE & 2.5172 & 13.4270 & 12.0416 & 1.8644 & 0.9180\\

 &  &  & GEE & 18.2916 & 20.8934 & 15.4059 & 7.7068 & 0.6860\\

 &  & \multirow[t]{-2}{*}{\raggedright\arraybackslash Exchangeable} & CWGEE & 2.5719 & 13.3070 & 12.0423 & 1.8351 & 0.9190\\

 &  &  & GEE & 17.4354 & 18.5425 & 13.4691 & 6.4747 & 0.6470\\

\multirow[t]{-18}{*}{\raggedleft\arraybackslash 30} & \multirow[t]{-6}{*}{\raggedright\arraybackslash Severe} & \multirow[t]{-2}{*}{\raggedright\arraybackslash AR1} & CWGEE & 3.4989 & 12.3243 & 10.9114 & 1.6398 & 0.9060\\
\cmidrule{1-9}
 &  &  & GEE & -18.6801 & 11.6607 & 9.0713 & 4.8478 & 0.4640\\

 &  & \multirow[t]{-2}{*}{\raggedright\arraybackslash Independent} & CWGEE & -0.1230 & 4.0919 & 3.8899 & 0.1674 & 0.9490\\

 &  &  & GEE & -18.6573 & 11.6242 & 9.0716 & 4.8308 & 0.4680\\

 &  & \multirow[t]{-2}{*}{\raggedright\arraybackslash Exchangeable} & CWGEE & -0.1202 & 4.0979 & 3.8899 & 0.1679 & 0.9510\\

 &  &  & GEE & -18.2699 & 11.0817 & 8.6352 & 4.5647 & 0.4300\\

 & \multirow[t]{-6}{*}{\raggedright\arraybackslash Healthy} & \multirow[t]{-2}{*}{\raggedright\arraybackslash AR1} & CWGEE & -0.0372 & 4.3523 & 4.1363 & 0.1893 & 0.9560\\

 &  &  & GEE & -4.9360 & 3.5559 & 3.1795 & 0.3700 & 0.6000\\

 &  & \multirow[t]{-2}{*}{\raggedright\arraybackslash Independent} & CWGEE & -0.7698 & 3.0653 & 2.9475 & 0.0998 & 0.9220\\

 &  &  & GEE & -4.8992 & 3.5410 & 3.1797 & 0.3653 & 0.6010\\

 &  & \multirow[t]{-2}{*}{\raggedright\arraybackslash Exchangeable} & CWGEE & -0.8105 & 3.0583 & 2.9475 & 0.1000 & 0.9200\\

 &  &  & GEE & -4.4513 & 3.4777 & 3.1639 & 0.3190 & 0.6400\\

 & \multirow[t]{-6}{*}{\raggedright\arraybackslash Moderate} & \multirow[t]{-2}{*}{\raggedright\arraybackslash AR1} & CWGEE & -1.3210 & 3.1083 & 3.0100 & 0.1140 & 0.9000\\

 &  &  & GEE & 23.6161 & 10.2092 & 8.6934 & 6.6184 & 0.2330\\

 &  & \multirow[t]{-2}{*}{\raggedright\arraybackslash Independent} & CWGEE & 0.8928 & 5.2258 & 4.8382 & 0.2808 & 0.9280\\

 &  &  & GEE & 23.5616 & 10.1741 & 8.6931 & 6.5856 & 0.2330\\

 &  & \multirow[t]{-2}{*}{\raggedright\arraybackslash Exchangeable} & CWGEE & 0.9152 & 5.2062 & 4.8382 & 0.2791 & 0.9280\\

 &  &  & GEE & 22.6788 & 9.6945 & 8.2263 & 6.0822 & 0.2190\\

\multirow[t]{-18}{*}{\raggedleft\arraybackslash 200} & \multirow[t]{-6}{*}{\raggedright\arraybackslash Severe} & \multirow[t]{-2}{*}{\raggedright\arraybackslash AR1} & CWGEE & 1.3879 & 5.1162 & 4.7305 & 0.2808 & 0.9300\\
\bottomrule
\multicolumn{9}{l}{Abbreviations: MCSD - Monte Carlo standard deviation, ASE - Average estimated standard error,}\\
\multicolumn{9}{l}{MSE - Mean squared error.}\\
\multicolumn{9}{l}{Bias, MCSD, ASE, and MSE are multiplied by $10^2$.}\\
\end{tabular}
}}
\end{table}
\FloatBarrier

\FloatBarrier
\begin{table}[ht]\tiny{
\caption{Simulation results for the pseudo-value regression parameter $\beta_1$ under the setting where the cluster sizes are non-informative. The pseudo-values were computed at 10 time points equally spread on the transition time scale. We consider three different correlation structures (independent, exchangeable, and AR1) for modeling the temporal correlation in the estimating equations.}
\label{web:bias_ICSm30_200_nics}
\centering
\resizebox{1.0\textwidth}{!}{%
\begin{tabular}[t]{lrllrllll}
\toprule
$m$ & State & Cor. Structure & Model & Bias & MCSD & ASE & MSE & Coverage\\
\midrule
 &  &  & GEE & -0.6531 & 8.2675 & 7.9622 & 0.6871 & 0.9340\\

 &  & \multirow[t]{-2}{*}{\raggedright\arraybackslash Independent} & CWGEE & -0.6476 & 8.2038 & 7.8698 & 0.6765 & 0.9390\\

 &  &  & GEE & -0.6556 & 8.2522 & 7.9623 & 0.6846 & 0.9340\\

 &  & \multirow[t]{-2}{*}{\raggedright\arraybackslash Exchangeable} & CWGEE & -0.6509 & 8.1892 & 7.8699 & 0.6742 & 0.9390\\

 &  &  & GEE & -0.6780 & 8.1926 & 7.8691 & 0.6751 & 0.9400\\

 & \multirow[t]{-6}{*}{\raggedright\arraybackslash Healthy} & \multirow[t]{-2}{*}{\raggedright\arraybackslash AR1} & CWGEE & -0.6863 & 8.1414 & 7.7945 & 0.6669 & 0.9390\\

 &  &  & GEE & 0.1830 & 4.2132 & 3.9437 & 0.1777 & 0.9260\\

 &  & \multirow[t]{-2}{*}{\raggedright\arraybackslash Independent} & CWGEE & 0.2034 & 4.2049 & 3.9557 & 0.1770 & 0.9280\\

 &  &  & GEE & 0.1480 & 4.1876 & 3.9438 & 0.1754 & 0.9280\\

 &  & \multirow[t]{-2}{*}{\raggedright\arraybackslash Exchangeable} & CWGEE & 0.1689 & 4.1806 & 3.9557 & 0.1749 & 0.9290\\

 &  &  & GEE & -0.4470 & 4.0777 & 3.7842 & 0.1681 & 0.9210\\

 & \multirow[t]{-6}{*}{\raggedright\arraybackslash Moderate} & \multirow[t]{-2}{*}{\raggedright\arraybackslash AR1} & CWGEE & -0.4174 & 4.0990 & 3.8249 & 0.1696 & 0.9230\\

 &  &  & GEE & 0.4702 & 10.8197 & 10.2812 & 1.1717 & 0.9330\\

 &  & \multirow[t]{-2}{*}{\raggedright\arraybackslash Independent} & CWGEE & 0.4442 & 10.6655 & 10.1408 & 1.1384 & 0.9350\\

 &  &  & GEE & 0.5104 & 10.7694 & 10.2814 & 1.1612 & 0.9340\\

 &  & \multirow[t]{-2}{*}{\raggedright\arraybackslash Exchangeable} & CWGEE & 0.4848 & 10.6169 & 10.1409 & 1.1284 & 0.9360\\

 &  &  & GEE & 1.1485 & 10.1874 & 9.7024 & 1.0500 & 0.9320\\

\multirow[t]{-18}{*}{\raggedleft\arraybackslash 30} & \multirow[t]{-6}{*}{\raggedright\arraybackslash Severe} & \multirow[t]{-2}{*}{\raggedright\arraybackslash AR1} & CWGEE & 1.1284 & 10.0671 & 9.5888 & 1.0252 & 0.9370\\
\cmidrule{1-9}
 &  &  & GEE & 0.0568 & 3.1996 & 3.1782 & 0.1023 & 0.9450\\

 &  & \multirow[t]{-2}{*}{\raggedright\arraybackslash Independent} & CWGEE & 0.0379 & 3.1843 & 3.1227 & 0.1013 & 0.9450\\

 &  &  & GEE & 0.0609 & 3.1971 & 3.1782 & 0.1021 & 0.9440\\

 &  & \multirow[t]{-2}{*}{\raggedright\arraybackslash Exchangeable} & CWGEE & 0.0419 & 3.1819 & 3.1227 & 0.1012 & 0.9450\\

 &  &  & GEE & 0.1237 & 3.1851 & 3.1709 & 0.1015 & 0.9500\\

 & \multirow[t]{-6}{*}{\raggedright\arraybackslash Healthy} & \multirow[t]{-2}{*}{\raggedright\arraybackslash AR1} & CWGEE & 0.1026 & 3.1731 & 3.1187 & 0.1007 & 0.9480\\

 &  &  & GEE & 0.1615 & 1.5318 & 1.5567 & 0.0237 & 0.9570\\

 &  & \multirow[t]{-2}{*}{\raggedright\arraybackslash Independent} & CWGEE & 0.1674 & 1.5303 & 1.5601 & 0.0237 & 0.9580\\

 &  &  & GEE & 0.1460 & 1.5300 & 1.5567 & 0.0236 & 0.9570\\

 &  & \multirow[t]{-2}{*}{\raggedright\arraybackslash Exchangeable} & CWGEE & 0.1520 & 1.5283 & 1.5601 & 0.0236 & 0.9590\\

 &  &  & GEE & -0.1103 & 1.5475 & 1.5630 & 0.0240 & 0.9530\\

 & \multirow[t]{-6}{*}{\raggedright\arraybackslash Moderate} & \multirow[t]{-2}{*}{\raggedright\arraybackslash AR1} & CWGEE & -0.1036 & 1.5457 & 1.5733 & 0.0240 & 0.9500\\

 &  &  & GEE & -0.2182 & 4.1171 & 4.0752 & 0.1698 & 0.9460\\

 &  & \multirow[t]{-2}{*}{\raggedright\arraybackslash Independent} & CWGEE & -0.2053 & 4.0854 & 4.0015 & 0.1672 & 0.9420\\

 &  &  & GEE & -0.2047 & 4.1109 & 4.0752 & 0.1692 & 0.9450\\

 &  & \multirow[t]{-2}{*}{\raggedright\arraybackslash Exchangeable} & CWGEE & -0.1917 & 4.0789 & 4.0015 & 0.1666 & 0.9420\\

 &  &  & GEE & 0.0017 & 4.0447 & 4.0014 & 0.1634 & 0.9440\\

\multirow[t]{-18}{*}{\raggedleft\arraybackslash 200} & \multirow[t]{-6}{*}{\raggedright\arraybackslash Severe} & \multirow[t]{-2}{*}{\raggedright\arraybackslash AR1} & CWGEE & 0.0161 & 4.0093 & 3.9333 & 0.1606 & 0.9420\\
\bottomrule
\multicolumn{9}{l}{Abbreviations: MCSD - Monte Carlo standard deviation, ASE - Average estimated standard error,}\\
\multicolumn{9}{l}{MSE - Mean squared error.}\\
\multicolumn{9}{l}{Bias, MCSD, ASE, and MSE are multiplied by $10^2$.}\\
\end{tabular}
}}
\end{table}
\FloatBarrier

\section{Sensitivity to choice of $r$} \label{websec:rsens}

\begin{table}[H]
\centering 
\scriptsize{
\caption{\small{Simulation results for the pseudo-value regression parameter $\beta_1$ under the setting where the cluster sizes are informative. The results are based on the CWGEE model for the healthy state. The pseudo-values were computed at $r$ time points equally spread across the transition time scale.}} 
\label{webtab:addsim1} 
\begin{tabular}[t]{lrrrllll}
\toprule
Cor. Structure & $m$ & $r$ & Bias & MCSD & ASE & MSE & Coverage\\
\midrule
 &  & 5 & -1.0900 & 10.8673 & 10.2507 & 1.1917 & 0.9380\\

 &  & 7 & -0.9996 & 10.6832 & 10.1136 & 1.1502 & 0.9390\\

 &  & 10 & -0.9935 & 10.6866 & 10.1175 & 1.1508 & 0.9400\\

 &  & 15 & -0.8837 & 10.8117 & 10.2171 & 1.1756 & 0.9340\\

 & \multirow[t]{-5}{*}{\raggedleft\arraybackslash 30} & 20 & -0.8911 & 10.7506 & 10.1576 & 1.1625 & 0.9340\\
\cmidrule{2-8}

 &  & 5 & -0.4982 & 4.1181 & 4.0521 & 0.1719 & 0.9580\\

 &  & 7 & -0.4592 & 3.9966 & 3.9242 & 0.1617 & 0.9570\\

 &  & 10 & -0.4909 & 3.9984 & 3.9111 & 0.1621 & 0.9620\\

 &  & 15 & -0.4491 & 4.0211 & 3.9110 & 0.1636 & 0.9550\\

\multirow[t]{-10}{*}{\raggedright\arraybackslash Independent} & \multirow[t]{-5}{*}{\raggedleft\arraybackslash 200} & 20 & -0.4711 & 3.9960 & 3.8912 & 0.1617 & 0.9590\\
\cmidrule{1-8}
 &  & 5 & -1.1238 & 10.8767 & 10.2507 & 1.1945 & 0.9380\\

 &  & 7 & -1.0280 & 10.6798 & 10.1137 & 1.1500 & 0.9370\\

 &  & 10 & -1.0138 & 10.6736 & 10.1175 & 1.1484 & 0.9380\\

 &  & 15 & -0.8996 & 10.7903 & 10.2171 & 1.1712 & 0.9350\\

 & \multirow[t]{-5}{*}{\raggedleft\arraybackslash 30} & 20 & -0.9010 & 10.7345 & 10.1577 & 1.1593 & 0.9350\\
\cmidrule{2-8}

 &  & 5 & -0.4958 & 4.1158 & 4.0521 & 0.1717 & 0.9580\\

 &  & 7 & -0.4552 & 3.9996 & 3.9242 & 0.1619 & 0.9570\\

 &  & 10 & -0.4866 & 3.9983 & 3.9111 & 0.1621 & 0.9600\\

 &  & 15 & -0.4470 & 4.0175 & 3.9110 & 0.1632 & 0.9550\\

\multirow[t]{-10}{*}{\raggedright\arraybackslash Exchangeable} & \multirow[t]{-5}{*}{\raggedleft\arraybackslash 200} & 20 & -0.4687 & 3.9928 & 3.8912 & 0.1615 & 0.9590\\
\cmidrule{1-8}
 &  & 5 & -1.2846 & 11.1620 & 10.4360 & 1.2612 & 0.9260\\

 &  & 7 & -1.2538 & 11.1777 & 10.4565 & 1.2639 & 0.9290\\

 &  & 10 & -1.3237 & 11.4641 & 10.7023 & 1.3305 & 0.9340\\

 &  & 15 & -1.3695 & 11.7404 & 10.9430 & 1.3958 & 0.9360\\

 & \multirow[t]{-5}{*}{\raggedleft\arraybackslash 30} & 20 & -1.3733 & 12.1551 & 11.2935 & 1.4949 & 0.9360\\
\cmidrule{2-8}

 &  & 5 & -0.4880 & 4.1465 & 4.0906 & 0.1741 & 0.9590\\

 &  & 7 & -0.4199 & 4.0962 & 4.0525 & 0.1694 & 0.9550\\

 &  & 10 & -0.4100 & 4.1979 & 4.1670 & 0.1777 & 0.9520\\

 &  & 15 & -0.3737 & 4.3240 & 4.3026 & 0.1882 & 0.9560\\

\multirow[t]{-10}{*}{\raggedright\arraybackslash AR1} & \multirow[t]{-5}{*}{\raggedleft\arraybackslash 200} & 20 & -0.3186 & 4.5261 & 4.5247 & 0.2057 & 0.9500\\
\bottomrule
\multicolumn{8}{l}{Abbreviations: $r$ - Number of time points,}\\
\multicolumn{8}{l}{MCSD - Monte Carlo standard deviation,}\\
\multicolumn{8}{l}{ASE - Average estimated standard error,}\\
\multicolumn{8}{l}{MSE - Mean squared error.}\\
\multicolumn{8}{l}{Bias, MCSD, ASE, and MSE are multiplied by $10^2$.}\\
\end{tabular}}
\end{table}

\newpage

\section{Multilevel design}\label{websec:ICG}
In Section 3.3 of the main paper, we introduced informative intra-cluster group (ICG) sizes. In such scenarios, there is an additional hierarchy due to the presence of known groups within the distinct clusters, and where the size of the known groups may be related to the observed outcomes. Let $i=1,\ldots,m$ index the clusters and let $j=1,\ldots,n_i$ index the cluster units. Suppose each cluster comprises two known groups with at least one observation in each group. The methods described here can be easily modified to the cases where there are $> 2$ known groups within a given cluster, and certain clusters may not have units belonging to any of the known groups. Suppose $G_{ij}$ is a subject-level binary level covariate indicating group membership, while $n_{i0}$ and $n_{i1}$ denote the respective group sizes corresponding to groups $G_{i\cdot}=0$ and $G_{i\cdot}=1$ in cluster $i$. Under this data structure, we now proceed with implementation of the pseudo-value regression for testing the hypothesis $H_0: P\{X_{ij}(t) = \ell | G_{ij} = 0, \mathbf{Z}_{ij}=\mathbf{z}\} = P\{X_{ij}(t) = \ell | G_{ij} = 1, \mathbf{Z}_{ij}=\mathbf{z}\}$ for $\ell \in \mathcal{S}$ and $t \geq 0$. 

We have shown that correct inference based on the pseudo-value approach is driven by formulating the estimating equations with appropriate weights. For a marginal inference, we consider three possible weights to adjust for the informative ICG sizes: (i) $w_{ij}=1$, (ii) $w_{ij}=\frac{1}{n_{i}}$, and (iii)  $w_{ij}=\frac{1}{2n_{iG_{ij}}}$, where $n_i$ is the number of units in cluster $i$, and $n_{iG_{ij}}$ denotes the number of units in cluster $i$ with the same group membership as unit $j$. The utilization of the weights in the estimating equations respectively correspond to the traditional GEE \cite{liang1986b}, the CWGEE \cite{williamson2003b}, and the doubly-weighted estimating equation (DWGEE)\citep{huang2011b}. Using simulation studies, we now investigate the appropriateness of these estimating equations based on pseudo-value responses for the analysis of clustered CS data with ICS.

Following the lognormal model given by Equation 9 in Section 4 of the main body of the paper, we now simulate clustered exit times from state 1. However, in this case, $Z_{1,ij} = G_{ij}$ is a subject-level binary covariate indicating group membership. We follow the simulation design similar to that presented in Section 4. The true exit times from state 1 are simulated using the Lognormal AFT model, given by 

\begin{align} 
    \log(T_{1 \cdot,ij}) = \delta_1 G_{ij} + \delta_2 Z_{2,ij} +  \nu_i + \sigma\varepsilon_{ij}
\end{align} 
where, $G_{ij}$ is a subject-level grouping indicator, $Z_{2,ij} \sim \text{N}(0, 0.15)$ is a subject-level continuous covariate, $\nu_i \sim \text{N}(0, 0.25)$ is a cluster-level random effect, and $\varepsilon_{ij} \sim \text{N}(0,\ 1)$ is the subject-level random error. We simulate $n_{i0} \sim \text{Poisson}\{\exp(3.2 + 2.5\nu_i)\}+2$ and $n_{i1} \sim \text{Poisson}\{\exp(0.8 + 1.5\nu_i)\}+2$. Except stated otherwise, we follow the simulation designs presented in Section 4 of the main body of the paper.

\subsection{Simulation results}
The simulation results are described in Section Section 4 (last paragraph) of the main paper. The corresponding Tables follow. 

\newpage 

\begin{table}[hbt!]\centering\scriptsize{
\vspace{-1.3cm}
\caption{\small{Empirical size (and 95\% confidence intervals), and power comparisons of hypothesis tests corresponding to the GEE and CWGEE models (nominal $\alpha = 0.05$) under where the intra-cluster group sizes are informative. The pseudo-values were computed at 10 time points equally spread on the transition time scale. We consider three different correlation structures (independent, exchangeable, and AR1) for modeling the temporal correlation in the estimating equations.}}\label{webtab:pow_sbics}
\resizebox{0.8\textwidth}{!}{%
\begin{tabular}[t]{rlllllll}
\toprule
\multicolumn{5}{c}{ } & \multicolumn{3}{c}{Power(under $\delta_1 > 0$)} \\
\cmidrule(l{3pt}r{3pt}){6-8}
$m$ & State & Cor. Structure & Model & Size (CI) & 0.25 & 0.50 & 0.75\\
\midrule
 &  &  & GEE & 0.3590 (0.3293, 0.3887) & 0.1780 & 0.7520 & 0.9730\\

 &  &  & CWGEE & 0.3010 (0.2726, 0.3294) & 0.1330 & 0.7700 & 0.9860\\

 &  & \multirow[t]{-3}{*}{\raggedright\arraybackslash Independent} & DWGEE & 0.0700 (0.0542, 0.0858) & 0.6890 & 0.9580 & 0.9960\\

 &  &  & GEE & 0.3590 (0.3293, 0.3887) & 0.1790 & 0.7510 & 0.9720\\

 &  &  & CWGEE & 0.3050 (0.2765, 0.3335) & 0.1360 & 0.7690 & 0.9860\\

 &  & \multirow[t]{-3}{*}{\raggedright\arraybackslash Exchangeable} & DWGEE & 0.0720 (0.0560, 0.0880) & 0.6830 & 0.9570 & 0.9960\\

 &  &  & GEE & 0.3190 (0.2901, 0.3479) & 0.1460 & 0.7200 & 0.9640\\

 &  &  & CWGEE & 0.2660 (0.2386, 0.2934) & 0.1250 & 0.7170 & 0.9740\\

 & \multirow[t]{-9}{*}{\raggedright\arraybackslash Healthy} & \multirow[t]{-3}{*}{\raggedright\arraybackslash AR1} & DWGEE & 0.0600 (0.0453, 0.0747) & 0.5680 & 0.9350 & 0.9910\\

 &  &  & GEE & 0.1340 (0.1129, 0.1551) & 0.0630 & 0.1090 & 0.1080\\

 &  &  & CWGEE & 0.1360 (0.1148, 0.1572) & 0.0670 & 0.1390 & 0.2090\\

 &  & \multirow[t]{-3}{*}{\raggedright\arraybackslash Independent} & DWGEE & 0.0760 (0.0596, 0.0924) & 0.1100 & 0.2040 & 0.2280\\

 &  &  & GEE & 0.1300 (0.1092, 0.1508) & 0.0630 & 0.1080 & 0.1030\\

 &  &  & CWGEE & 0.1370 (0.1157, 0.1583) & 0.0660 & 0.1340 & 0.2060\\

 &  & \multirow[t]{-3}{*}{\raggedright\arraybackslash Exchangeable} & DWGEE & 0.0770 (0.0605, 0.0935) & 0.1080 & 0.1990 & 0.2260\\

 &  &  & GEE & 0.1100 (0.0906, 0.1294) & 0.0600 & 0.0890 & 0.0810\\

 &  &  & CWGEE & 0.1190 (0.0989, 0.1391) & 0.0540 & 0.1040 & 0.1300\\

 & \multirow[t]{-9}{*}{\raggedright\arraybackslash Moderate} & \multirow[t]{-3}{*}{\raggedright\arraybackslash AR1} & DWGEE & 0.0900 (0.0723, 0.1077) & 0.0680 & 0.1200 & 0.1460\\

 &  &  & GEE & 0.4110 (0.3805, 0.4415) & 0.1550 & 0.7400 & 0.9770\\

 &  &  & CWGEE & 0.3400 (0.3106, 0.3694) & 0.1460 & 0.7800 & 0.9860\\

 &  & \multirow[t]{-3}{*}{\raggedright\arraybackslash Independent} & DWGEE & 0.0610 (0.0462, 0.0758) & 0.7310 & 0.9520 & 0.9950\\

 &  &  & GEE & 0.4060 (0.3756, 0.4364) & 0.1550 & 0.7370 & 0.9770\\

 &  &  & CWGEE & 0.3330 (0.3038, 0.3622) & 0.1470 & 0.7800 & 0.9860\\

 &  & \multirow[t]{-3}{*}{\raggedright\arraybackslash Exchangeable} & DWGEE & 0.0580 (0.0435, 0.0725) & 0.7300 & 0.9500 & 0.9950\\

 &  &  & GEE & 0.3300 (0.3009, 0.3591) & 0.1560 & 0.6980 & 0.9640\\

 &  &  & CWGEE & 0.2600 (0.2328, 0.2872) & 0.1570 & 0.7190 & 0.9790\\

\multirow[t]{-27}{*}{\raggedleft\arraybackslash 30} & \multirow[t]{-9}{*}{\raggedright\arraybackslash Severe} & \multirow[t]{-3}{*}{\raggedright\arraybackslash AR1} & DWGEE & 0.0600 (0.0453, 0.0747) & 0.6420 & 0.9400 & 0.9920\\
\cmidrule{1-8}
 &  &  & GEE & 0.9820 (0.9738, 0.9902) & 0.3480 & 1.0000 & 1.0000\\

 &  &  & CWGEE & 0.9380 (0.9231, 0.9529) & 0.4670 & 0.9980 & 1.0000\\

 &  & \multirow[t]{-3}{*}{\raggedright\arraybackslash Independent} & DWGEE & 0.0500 (0.0365, 0.0635) & 0.9960 & 1.0000 & 1.0000\\

 &  &  & GEE & 0.9820 (0.9738, 0.9902) & 0.3450 & 1.0000 & 1.0000\\

 &  &  & CWGEE & 0.9390 (0.9242, 0.9538) & 0.4640 & 0.9980 & 1.0000\\

 &  & \multirow[t]{-3}{*}{\raggedright\arraybackslash Exchangeable} & DWGEE & 0.0500 (0.0365, 0.0635) & 0.9960 & 1.0000 & 1.0000\\

 &  &  & GEE & 0.9750 (0.9653, 0.9847) & 0.3300 & 0.9990 & 1.0000\\

 &  &  & CWGEE & 0.9030 (0.8847, 0.9213) & 0.4180 & 0.9970 & 0.9990\\

 & \multirow[t]{-9}{*}{\raggedright\arraybackslash Healthy} & \multirow[t]{-3}{*}{\raggedright\arraybackslash AR1} & DWGEE & 0.0520 (0.0382, 0.0658) & 0.9920 & 1.0000 & 0.9990\\

 &  &  & GEE & 0.3570 (0.3273, 0.3867) & 0.0720 & 0.3310 & 0.3390\\

 &  &  & CWGEE & 0.4350 (0.4043, 0.4657) & 0.1020 & 0.6590 & 0.8580\\

 &  & \multirow[t]{-3}{*}{\raggedright\arraybackslash Independent} & DWGEE & 0.0450 (0.0322, 0.0578) & 0.5230 & 0.9020 & 0.9020\\

 &  &  & GEE & 0.3510 (0.3214, 0.3806) & 0.0690 & 0.3250 & 0.3360\\

 &  &  & CWGEE & 0.4300 (0.3993, 0.4607) & 0.1040 & 0.6560 & 0.8560\\

 &  & \multirow[t]{-3}{*}{\raggedright\arraybackslash Exchangeable} & DWGEE & 0.0460 (0.0330, 0.0590) & 0.5180 & 0.8940 & 0.8960\\

 &  &  & GEE & 0.2790 (0.2512, 0.3068) & 0.0650 & 0.2510 & 0.2670\\

 &  &  & CWGEE & 0.3360 (0.3067, 0.3653) & 0.0910 & 0.5530 & 0.7390\\

 & \multirow[t]{-9}{*}{\raggedright\arraybackslash Moderate} & \multirow[t]{-3}{*}{\raggedright\arraybackslash AR1} & DWGEE & 0.0560 (0.0417, 0.0703) & 0.3890 & 0.7690 & 0.7910\\

 &  &  & GEE & 0.9900 (0.9838, 0.9962) & 0.3680 & 1.0000 & 1.0000\\

 &  &  & CWGEE & 0.9510 (0.9376, 0.9644) & 0.4750 & 0.9980 & 0.9990\\

 &  & \multirow[t]{-3}{*}{\raggedright\arraybackslash Independent} & DWGEE & 0.0380 (0.0261, 0.0499) & 0.9950 & 1.0000 & 1.0000\\

 &  &  & GEE & 0.9900 (0.9838, 0.9962) & 0.3640 & 1.0000 & 1.0000\\

 &  &  & CWGEE & 0.9520 (0.9388, 0.9652) & 0.4750 & 0.9980 & 0.9990\\

 &  & \multirow[t]{-3}{*}{\raggedright\arraybackslash Exchangeable} & DWGEE & 0.0380 (0.0261, 0.0499) & 0.9950 & 1.0000 & 1.0000\\

 &  &  & GEE & 0.9820 (0.9738, 0.9902) & 0.3410 & 1.0000 & 1.0000\\

 &  &  & CWGEE & 0.9310 (0.9153, 0.9467) & 0.4430 & 0.9960 & 0.9990\\

\multirow[t]{-27}{*}{\raggedleft\arraybackslash 200} & \multirow[t]{-9}{*}{\raggedright\arraybackslash Severe} & \multirow[t]{-3}{*}{\raggedright\arraybackslash AR1} & DWGEE & 0.0390 (0.0270, 0.0510) & 0.9920 & 0.9980 & 0.9990\\
\bottomrule
\end{tabular}}}
\end{table}

\newpage 

\begin{table}[hbt!]\centering\scriptsize{
\caption{\small{Simulation results for the pseudo-value regression parameter $\beta_1$ under the setting where the intra-cluster group sizes are informative. The pseudo-values were computed at 10 time points equally spread on the transition time scale. We consider three different correlation structures (independent, exchangeable, and AR1) for modeling the temporal correlation in the estimating equations.}}
\label{webtab:bias_ICSm30_200_sbics}
\begin{tabular}[t]{lrllrllll}
\toprule
$m$ & State & Correlation structure & Model & Bias & MCSD & ASE & MSE & Coverage\\
\midrule
 &  &  & GEE & -6.0039 & 5.2660 & 4.8178 & 0.6375 & 0.7550\\

 &  &  & CWGEE & -6.3896 & 4.3937 & 4.1129 & 0.6011 & 0.6250\\

 &  & \multirow[t]{-3}{*}{\raggedright\arraybackslash Independent} & DWGEE & -0.2668 & 4.1326 & 3.8870 & 0.1713 & 0.9290\\

 &  &  & GEE & -6.0133 & 5.2527 & 4.8182 & 0.6372 & 0.7520\\

 &  &  & CWGEE & -6.3857 & 4.4079 & 4.1128 & 0.6019 & 0.6200\\

 &  & \multirow[t]{-3}{*}{\raggedright\arraybackslash Exchangeable} & DWGEE & -0.2681 & 4.1407 & 3.8870 & 0.1720 & 0.9290\\

 &  &  & GEE & -6.1794 & 5.4881 & 5.0595 & 0.6827 & 0.7590\\

 &  &  & CWGEE & -6.3297 & 5.0943 & 4.7362 & 0.6599 & 0.7060\\

 & \multirow[t]{-9}{*}{\raggedright\arraybackslash Healthy} & \multirow[t]{-3}{*}{\raggedright\arraybackslash AR1} & DWGEE & -0.2806 & 4.8152 & 4.4836 & 0.2324 & 0.9350\\

 &  &  & GEE & -2.4973 & 4.2354 & 4.1223 & 0.2416 & 0.8780\\

 &  &  & CWGEE & -2.2022 & 3.8486 & 3.7520 & 0.1965 & 0.8830\\

 &  & \multirow[t]{-3}{*}{\raggedright\arraybackslash Independent} & DWGEE & -0.2366 & 4.0256 & 3.9858 & 0.1624 & 0.9340\\

 &  &  & GEE & -2.5042 & 4.2214 & 4.1224 & 0.2407 & 0.8810\\

 &  &  & CWGEE & -2.2109 & 3.8364 & 3.7520 & 0.1959 & 0.8820\\

 &  & \multirow[t]{-3}{*}{\raggedright\arraybackslash Exchangeable} & DWGEE & -0.2733 & 4.0155 & 3.9858 & 0.1618 & 0.9320\\

 &  &  & GEE & -2.6172 & 4.4313 & 4.2997 & 0.2647 & 0.8580\\

 &  &  & CWGEE & -2.3506 & 4.0463 & 3.9891 & 0.2188 & 0.8700\\

 & \multirow[t]{-9}{*}{\raggedright\arraybackslash Moderate} & \multirow[t]{-3}{*}{\raggedright\arraybackslash AR1} & DWGEE & -0.8517 & 4.2970 & 4.2443 & 0.1917 & 0.9090\\

 &  &  & GEE & 8.5012 & 6.4534 & 5.9374 & 1.1388 & 0.7100\\

 &  &  & CWGEE & 8.5918 & 6.0756 & 5.5082 & 1.1070 & 0.6380\\

 &  & \multirow[t]{-3}{*}{\raggedright\arraybackslash Independent} & DWGEE & 0.5034 & 5.4258 & 4.9616 & 0.2966 & 0.9350\\

 &  &  & GEE & 8.5195 & 6.4126 & 5.9378 & 1.1366 & 0.7060\\

 &  &  & CWGEE & 8.5979 & 6.0469 & 5.5081 & 1.1045 & 0.6450\\

 &  & \multirow[t]{-3}{*}{\raggedright\arraybackslash Exchangeable} & DWGEE & 0.5424 & 5.4003 & 4.9617 & 0.2943 & 0.9360\\

 &  &  & GEE & 8.8107 & 6.3096 & 5.8749 & 1.1740 & 0.6640\\

 &  &  & CWGEE & 8.6904 & 6.1798 & 5.6325 & 1.1367 & 0.6410\\

\multirow[t]{-27}{*}{\raggedleft\arraybackslash 30} & \multirow[t]{-9}{*}{\raggedright\arraybackslash Severe} & \multirow[t]{-3}{*}{\raggedright\arraybackslash AR1} & DWGEE & 1.1513 & 5.7101 & 5.2154 & 0.3390 & 0.9380\\
\cmidrule{1-9}
 &  &  & GEE & -6.4325 & 2.0868 & 2.0449 & 0.4573 & 0.0900\\

 &  &  & CWGEE & -6.5396 & 1.6324 & 1.5673 & 0.4543 & 0.0210\\

 &  & \multirow[t]{-3}{*}{\raggedright\arraybackslash Independent} & DWGEE & -0.0030 & 1.4925 & 1.4633 & 0.0223 & 0.9470\\

 &  &  & GEE & -6.4346 & 2.0876 & 2.0449 & 0.4576 & 0.0900\\

 &  &  & CWGEE & -6.5347 & 1.6450 & 1.5673 & 0.4541 & 0.0200\\

 &  & \multirow[t]{-3}{*}{\raggedright\arraybackslash Exchangeable} & DWGEE & 0.0016 & 1.5017 & 1.4633 & 0.0225 & 0.9470\\

 &  &  & GEE & -6.4669 & 2.1702 & 2.1023 & 0.4653 & 0.1070\\

 &  &  & CWGEE & -6.4576 & 1.8867 & 1.7456 & 0.4526 & 0.0430\\

 & \multirow[t]{-9}{*}{\raggedright\arraybackslash Healthy} & \multirow[t]{-3}{*}{\raggedright\arraybackslash AR1} & DWGEE & 0.0768 & 1.7214 & 1.6410 & 0.0297 & 0.9520\\

 &  &  & GEE & -2.3959 & 1.6673 & 1.6428 & 0.0852 & 0.6770\\

 &  &  & CWGEE & -2.0724 & 1.5088 & 1.4853 & 0.0657 & 0.6930\\

 &  & \multirow[t]{-3}{*}{\raggedright\arraybackslash Independent} & DWGEE & 0.0053 & 1.6104 & 1.5634 & 0.0259 & 0.9470\\

 &  &  & GEE & -2.3999 & 1.6686 & 1.6428 & 0.0854 & 0.6740\\

 &  &  & CWGEE & -2.0775 & 1.5112 & 1.4853 & 0.0660 & 0.6920\\

 &  & \multirow[t]{-3}{*}{\raggedright\arraybackslash Exchangeable} & DWGEE & -0.0114 & 1.6124 & 1.5634 & 0.0260 & 0.9460\\

 &  &  & GEE & -2.4678 & 1.7642 & 1.7388 & 0.0920 & 0.6750\\

 &  &  & CWGEE & -2.1617 & 1.6181 & 1.5977 & 0.0729 & 0.6920\\

 & \multirow[t]{-9}{*}{\raggedright\arraybackslash Moderate} & \multirow[t]{-3}{*}{\raggedright\arraybackslash AR1} & DWGEE & -0.2878 & 1.7207 & 1.6836 & 0.0304 & 0.9370\\

 &  &  & GEE & 8.8284 & 2.4921 & 2.4233 & 0.8415 & 0.0470\\

 &  &  & CWGEE & 8.6121 & 2.2217 & 2.1015 & 0.7910 & 0.0170\\

 &  & \multirow[t]{-3}{*}{\raggedright\arraybackslash Independent} & DWGEE & -0.0024 & 1.9386 & 1.8379 & 0.0375 & 0.9520\\

 &  &  & GEE & 8.8347 & 2.4885 & 2.4233 & 0.8424 & 0.0450\\

 &  &  & CWGEE & 8.6122 & 2.2279 & 2.1014 & 0.7913 & 0.0170\\

 &  & \multirow[t]{-3}{*}{\raggedright\arraybackslash Exchangeable} & DWGEE & 0.0106 & 1.9432 & 1.8379 & 0.0377 & 0.9540\\

 &  &  & GEE & 8.9359 & 2.5263 & 2.4645 & 0.8623 & 0.0450\\

 &  &  & CWGEE & 8.6170 & 2.3877 & 2.2196 & 0.7995 & 0.0340\\

\multirow[t]{-27}{*}{\raggedleft\arraybackslash 200} & \multirow[t]{-9}{*}{\raggedright\arraybackslash Severe} & \multirow[t]{-3}{*}{\raggedright\arraybackslash AR1} & DWGEE & 0.2157 & 2.1182 & 1.9935 & 0.0453 & 0.9550\\
\bottomrule
\multicolumn{9}{l}{Abbreviations: MCSD - Monte Carlo standard deviation, ASE - Average estimated standard error,}\\
\multicolumn{9}{l}{MSE - Mean squared error.}\\
\multicolumn{9}{l}{Bias, MCSD, ASE, and MSE are multiplied by $10^2$.}\\
\end{tabular}
}
\end{table}

\newpage
\bibliography{Supplementary.bib}